\documentclass{article}
\PassOptionsToPackage{numbers}{natbib}
\usepackage[final]{neurips_2025}
\usepackage{graphicx}
\usepackage{svg}
\usepackage{etoolbox}
\usepackage{amsmath}
\usepackage{wrapfig}
\usepackage{caption}
\usepackage{multirow}
\usepackage{amssymb}
\usepackage{pifont}
\usepackage{tabularx}
\usepackage{makecell}
\usepackage{threeparttable}
\usepackage[dvipsnames]{xcolor}
\usepackage{authblk}

\newcommand{\cmark}{\textcolor{green!60!black}{\ding{51}}}
\newcommand{\xmark}{\textcolor{red!60!black}{\ding{55}}}

\usepackage[utf8]{inputenc} 
\usepackage[T1]{fontenc}    
\usepackage{hyperref}       
\usepackage{url}            
\usepackage{booktabs}       
\usepackage{amsfonts}       
\usepackage{nicefrac}       
\usepackage{microtype}      
\usepackage{xcolor}         
\definecolor{LinearProbeBlue}{RGB}{145,155,186}

\title{Brain Harmony: A Multimodal Foundation Model Unifying Morphology and Function into 1D Tokens}

\newcommand{\equalmark}{$^{*}$}
\newcommand{\corrmark}{$^{\dagger}$}

\author[1]{Zijian Dong\equalmark}
\author[1,2]{Ruilin Li\equalmark}
\author[1]{Joanna Su Xian Chong}
\author[1]{Niousha Dehestani}
\author[1]{Yinghui Teng}
\author[1]{Yi Lin}
\author[1]{Zhizhou Li}
\author[1]{Yichi Zhang}
\author[1]{Yapei Xie}
\author[1]{Leon Qi Rong Ooi}
\author[1]{B.T. Thomas Yeo}
\author[1]{Juan~Helen~Zhou\corrmark}

\affil[1]{National University of Singapore}
\affil[2]{Now at MiroMind AI}

\begin{document}

\maketitle
\begingroup
\renewcommand{\thefootnote}{}
\footnotetext{\equalmark Equal contribution}
\footnotetext{\corrmark Corresponding author: \texttt{helen.zhou@nus.edu.sg}}
\endgroup

\begin{abstract}
  We present \textbf{Brain Harmony  (BrainHarmonix)}, the first multimodal brain foundation model that unifies structural morphology and functional dynamics into compact 1D token representations. The model was pretrained on two of the largest neuroimaging datasets to date, encompassing 64,594 T1-weighted structural MRI 3D volumes (\textasciitilde 14 million images) and 70,933 functional MRI (fMRI) time series. BrainHarmonix is grounded in two foundational neuroscience principles: \emph{structure complements function} - structural and functional modalities offer distinct yet synergistic insights into brain organization; \emph{function follows structure} - brain functional dynamics are shaped by cortical morphology. The modular pretraining process involves single-modality training with geometric pre-alignment followed by modality fusion through shared brain hub tokens. Notably, our dynamics encoder uniquely handles fMRI time series with heterogeneous repetition times (TRs), addressing a major limitation in existing models. BrainHarmonix is also the first to deeply compress high-dimensional neuroimaging signals into unified, continuous 1D tokens, forming a compact latent space of the human brain. BrainHarmonix achieves strong generalization across diverse downstream tasks, including neurodevelopmental and neurodegenerative disorder classification and cognition prediction - consistently outperforming previous approaches. Our models - pretrained on 8 H100 GPUs - aim to catalyze a new era of AI-driven neuroscience powered by large-scale multimodal neuroimaging. Code is available at: \url{https://github.com/hzlab/Brain-Harmony}
\end{abstract}

\section{Introduction}
\label{intro}

\begin{figure}[!t]
    \centering
    \includegraphics[width=\columnwidth]{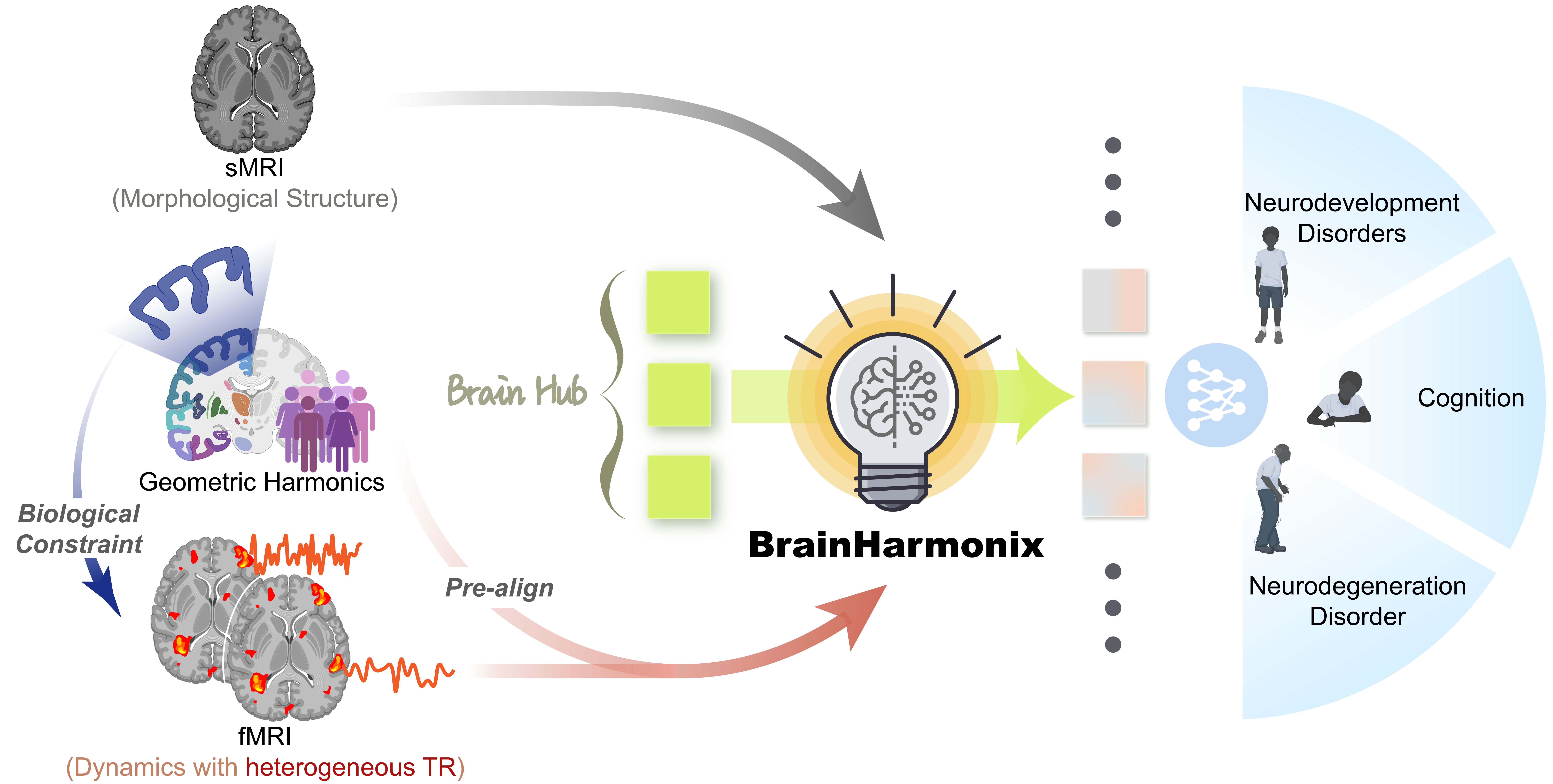}
    \caption{\textbf{Overview of Brain Harmony (BrainHarmonix).} Brain morphology from T1-weighted MRI (sMRI) and functional dynamics from fMRI are unified into compact 1D brain-hub tokens, which can be readily adapted to downstream tasks via an attached projection head. Specifically, functional dynamics are pre-aligned with group-level geometric harmonics, with built-in flexibility to handle heterogeneous repetition times (TRs). This fusion creates a compact yet expressive representation space that effectively captures the interplay between brain structure and function, supporting a broad range of downstream applications, including neurodevelopmental and neurodegenerative disorder classification and cognition prediction.}
    \label{fig1}
    \vspace{-25pt}
\end{figure}

The human brain is an extraordinarily complex organ, characterized by intricate anatomical architecture and dynamic functional processes. To investigate these aspects in vivo, researchers rely on neuroimaging techniques that probe brain structure and function. However, each neuroimaging modality captures only a single facet of this multifaceted system \cite{logothetis2008we,poldrack2015progress,calhoun2016multimodal}. This limitation highlights the necessity of multimodal neuroimaging approaches that combine complementary information (\emph{e.g.,} both structural and functional) to offer a holistic understanding of human cognition and improve clinical applications.

Recent advances in brain foundation models have transformed Artificial Intelligence (AI) for neuroimaging analysis from task-specific approaches to self-supervised pretrained models capable of adaptation across diverse downstream applications \cite{carobrainlm,dong2024brain,rui2024brainmvp,yang2024brainmass}. Despite their promising generalizability, these models focus on either brain structure \cite{rui2024brainmvp} (\emph{e.g.,} T1, T2-weighted MRI) or function \cite{carobrainlm,dong2024brain,yang2024brainmass} (\emph{e.g.,} functional MRI (fMRI)), without capturing the two complementary aspects simultaneously. Furthermore, recent neuroscience findings demonstrate that brain activity can be formulated as excitations of fundamental resonant modes shaped by the brain's geometry, revealing how morphological structure fundamentally constrains functional dynamics \cite{pang2023geometric}. Nevertheless, existing brain dynamics foundation models overlook this crucial constraint imposed by brain morphology. On the other hand, existing brain dynamics foundation models rely exclusively on fMRI datasets with homogeneous temporal resolutions \cite{carobrainlm,dong2024brain}, hindering the integration of datasets collected from diverse scanners and protocols with varying repetition times (TRs). Even within individual datasets or real-world clinical scenarios, multiple TRs often coexist \cite{adhd2012adhd,di2014autism,di2017enhancing}, rendering previous models infeasible for broader deployment. This limitation substantially reduces available sample sizes and constrains comprehensive modeling of brain dynamics across multiple temporal scales. While connectivity-based approaches are naturally agnostic to variations in TR \cite{yang2024brainmass}, they aggregate activity across entire scanning sessions, discarding essential non-stationary dynamics (\emph{e.g.}, transient state transitions and evolving co-activation patterns) in the blood-oxygen-level dependent (BOLD) signals \cite{chang2010time,calhoun2014chronnectome,carobrainlm,dong2024brain}.

Together, the aforementioned gaps highlight a fundamental challenge: creating comprehensive brain representations that effectively capture both structural and functional neuroimaging data with heterogeneous temporal resolutions. A critical step toward addressing this challenge is developing efficient methods to compress high-dimensional neuroimaging data into compact, information-dense representations. Transforming complex neuroimaging data into sequential 1D tokens offers a promising solution, potentially providing a unified framework for integrating multimodal information across diverse neuroimaging acquisitions.

In this paper, we propose \textbf{Brain Harmony (BrainHarmonix)} to address these critical gaps (Figure \ref{fig1}). Our major contributions include:
\textbf{(1)}~Developing the first multimodal brain foundation model to bridge morphological structure \emph{and} functional dynamics in a compact, information-rich representation space with 1D tokens.
\textbf{(2)}~Incorporating geometric harmonics to pre-align cortical morphology and functional organization, embedding structural constraints directly into functional representations. Imposing this population-level, physics-informed inductive bias can further enhance cross-subject and cross-dataset alignment.
\textbf{(3)}~Developing novel Temporal Adaptive Patch Embedding (TAPE) that enables scalable fMRI pretraining across heterogeneous TR values, overcoming a key limitation of existing models.
\textbf{(4)}~Introducing the first effective data augmentation for fMRI time series - downsampling to hierarchical TR levels - to accommodate heterogeneous TR distributions and enhance performance.
\textbf{(5)}~Finally, BrainHarmonix was benchmarked on a diverse set of downstream tasks, including the diagnosis of neurodevelopmental and neurodegenerative disorders, as well as the prediction of cognition. We demonstrate, for the first time, that \textcolor{BrickRed}{\emph{complex brain morphology and dynamics can be deeply compressed into unified continuous-valued 1D tokens}} that serve as holistic representations of the human brain.

\section{Related Work}

Recent brain foundation models have made significant advances in learning human brain representations. BrainLM \cite{carobrainlm} and Brain-JEPA \cite{dong2024brain} pioneered self-supervised learning for fMRI time series using masked prediction and joint-embedding approaches, respectively. While these models demonstrated promising generalizability through global representations, they suffer from two critical shortcomings: (1) they ignore brain structural information, and (2) due to their standard choice of patch embedding layer in transformers, cannot accommodate heterogeneous TRs common across - or even within - fMRI datasets. BDO \cite{park2025foundational} proposed a brain dynamics model based on stochastic optimal control, however, it focuses exclusively on brain dynamics, similar to Brain-JEPA and BrainLM. In addition, BrainMass \cite{yang2024brainmass} has been proposed as the first foundation model for brain functional connectivity and pretrained on diverse fMRI datasets. However, it focuses exclusively on static functional connectivity without capturing brain structural information or temporal dynamics. On the other hand, BrainMVP \cite{rui2024brainmvp} introduced self-supervised pretraining for 3D volumetric brain imaging that excels at learning correspondence among multi-parametric MRI, but fails to capture brain functional dynamics. This makes it suboptimal for gaining a comprehensive understanding of human brain functional organization, capturing individual differences in behavior, and detecting abnormal alterations associated with neuropsychiatric disorders. To the best of our knowledge, Brain Harmony (BrainHarmonix) addresses these limitations as \emph{the first multimodal foundation model that seamlessly integrates structural morphology with functional dynamics while accommodating variable TR values}. By unifying both modalities into 1D tokens, BrainHarmonix creates a compact and effective representational space that captures the holistic nature of the human brain.

\begin{figure}[!t]
    \centering
    \includegraphics[width=\columnwidth]{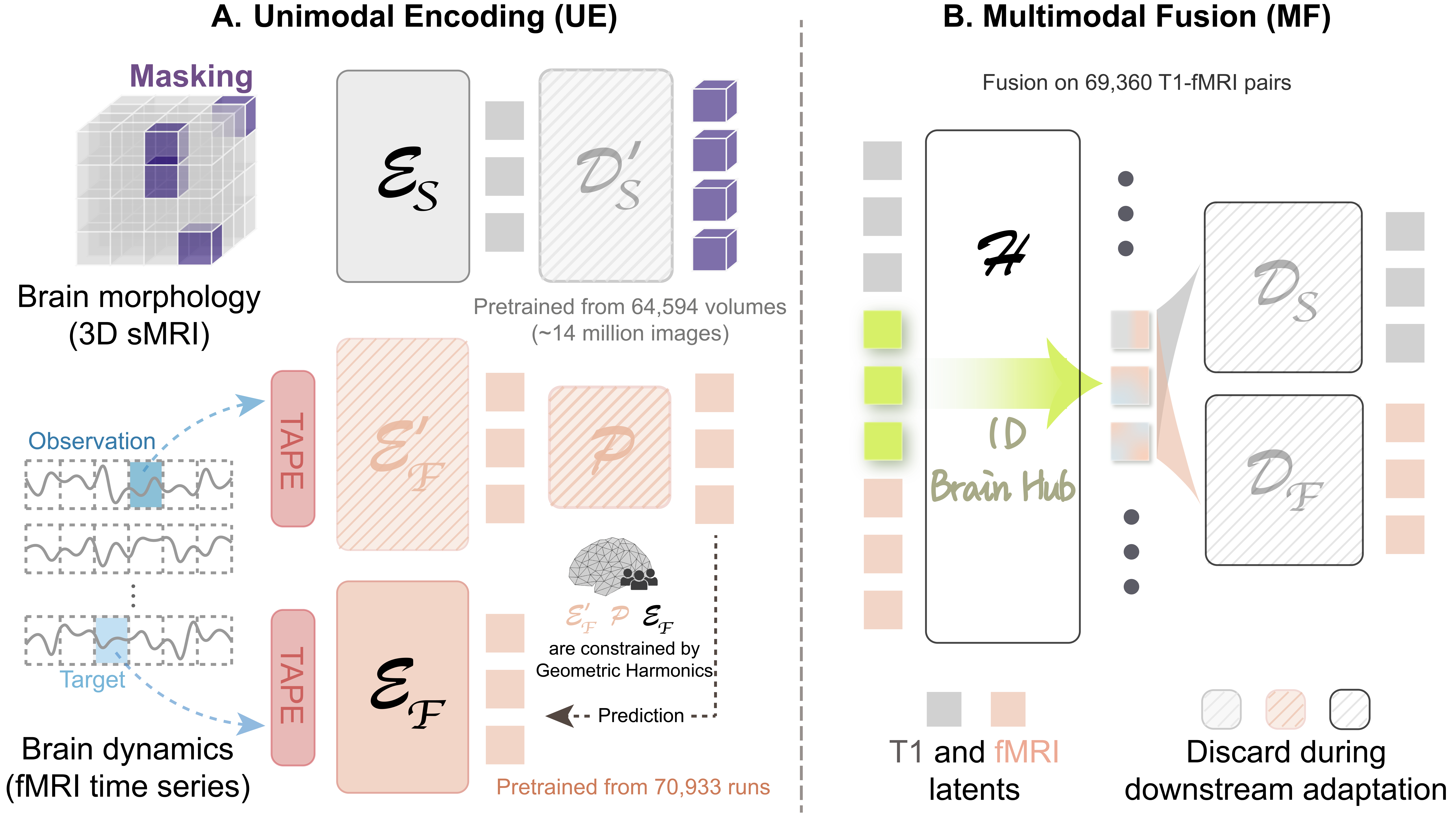}
    \caption{\textbf{Pretraining of Brain Harmony (BrainHarmonix).} \textbf{A. Unimodal Encoding (UE):} BrainHarmonix-S ($\varepsilon_{\scriptscriptstyle S}$) learns T1 structure via a Masked Autoencoder (MAE); gray cubes represent visible patches, while purple cubes are masked and reconstructed by the decoder ($\mathcal{D}'_{S}$). BrainHarmonix-F ($\varepsilon_{\scriptscriptstyle F}$) uses the Joint Embedding Predictive Architecture (JEPA) for fMRI, incorporating our Temporal Adaptive Patch Embedding (TAPE) for heterogeneous TRs and geometric harmonics for cortical alignment, with the observation encoder ($\varepsilon'_{\scriptscriptstyle F}$) and predictor ($\mathcal{P}$) following standard JEPA.
\textbf{B. Multimodal Fusion (MF):} The Harmonizer ($\mathcal{H}$) fuses structural and functional latents into 1D tokens (in green), then decoder ($\mathcal{D}_{S}$ \& $\mathcal{D}_{F}$) reconstruct modality-specific latents.}
    \label{fig2}
    \vspace{-10pt}
\end{figure}

\section{Method}

The pretraining of Brain Harmony (BrainHarmonix) comprises two sequential stages (Figure \ref{fig2}): \textbf{(1) Unimodal Encoding (UE)}: we first separately train modality-specific encoders for T1 (BrainHarmonix-S) and fMRI (BrainHarmonix-F). This separation allows flexible use of unpaired structural and functional data. For BrainHarmonix-S, we employ a 3D Masked Autoencoder (MAE) \cite{he2022masked} that effectively captures structural information from the \emph{largest} curation of T1 imaging datasets (given the widely adoption of MAE, readers are referred directly to Section \ref{imp} for implementation details). For BrainHarmonix-F, we propose two significant innovations to masked brain modeling: first, a geometric harmonics-based alignment method that pre-aligns brain dynamics with structural geometry; second, an innovative Temporal Adaptive Patch Embedding (TAPE) layer, enabling the encoder to flexibly accommodate \emph{any} TR for the first time.  Leveraging this unprecedented flexibility, we further introduce data augmentation techniques for fMRI time series, creating hierarchical TR values by downsampling high-resolution data. \textbf{(2) Multimodal Fusion (MF)}: modality-specific representations are fused through a set of learnable 1D brain hub tokens. These tokens act as a representational bottleneck, explicitly trained to reconstruct both structural and functional latents, resulting in a highly compact and unified latent space for human brain morphology and function.

\subsection{Unimodal Encoding (UE)}

In this subsection, we highlight two key innovations introduced in our approach for encoding fMRI dynamics (BrainHarmonix-F). First, we propose a geometric pre-alignment between brain dynamics and geometric harmonics, leveraging the foundational constraint that brain morphology inherently imposes on functional dynamics. Second, to effectively handle datasets with heterogeneous TRs, we introduce the Temporal Adaptive Patch Embedding (TAPE) layer, which enables token generation with consistent temporal length across varying TRs. For the encoding of T1 imaging (BrainHarmonix-S), readers are referred to the implementation details provided in Section \ref{imp}.

\subsubsection{Pre-alignment between brain dynamics and geometry}

\begin{minipage}[c]{0.5\textwidth}
    Recent neuroscientific research has revealed the profound relationship between brain morphology and functional dynamics, demonstrating that functional brain activity propagates as waves constrained by cortical geometry \cite{pang2023geometric}. In BrainHarmonix-F, we propose to pre-align brain dynamics with morphology. Specifically, we position brain ROIs in the transformer using geometric harmonics derived from population-level cortical surface mesh, thereby enhancing the structure-function coherence of the learned latent space of brain dynamics. 
\end{minipage}\hfill
\begin{minipage}[c]{0.45\textwidth}
    \centering
    \includegraphics[width=\linewidth]{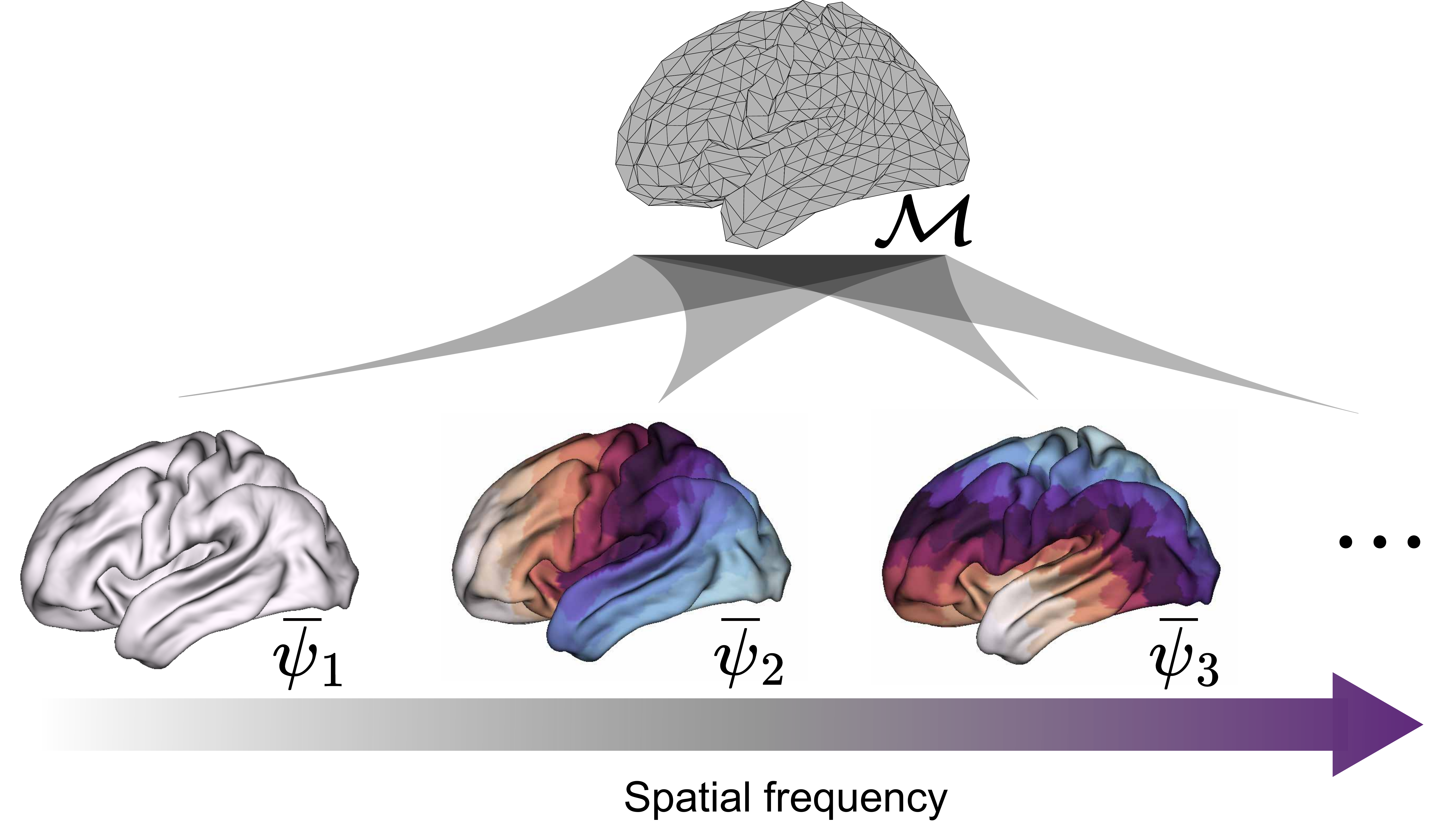}
    \vspace{-18pt}
    \captionsetup{hypcap=false}
    \captionof{figure}{\textbf{Geometric harmonics.}}
    \label{fig3} 
    \vspace{-5pt}
\end{minipage}

Geometric harmonics are the natural, orthogonal vibration patterns of the brain's folded surface. Given a mesh representation $\mathcal{M}$  of a population-averaged cortical surface derived from T1 imaging, the Laplace-Beltrami operator (LBO) $\Delta_{\mathcal M}$ is constructed to capture local vertex-to-vertex spatial relationships and cortical curvature. The corresponding eigenvalue problem can then be solved as follows:  
\vspace{-0.003cm}
\begin{equation}
\Delta_{\mathcal M}\,\psi \;=\; -\,\lambda\,\psi, \; \psi_i \xrightarrow{\;\downarrow \;} \overline{\psi}_i
\label{eq1}
\end{equation}
\vspace{-0.001cm}
where $\psi=\{\psi_1, \psi_2 ... \psi_i...\}$ is the sequence of geometric harmonics with the corresponding eigenvalues $\lambda=\{\lambda_1,\lambda_2...\lambda_i...\}$ ordered regarding spatial frequency (Figure \ref{fig3}). Each $\psi_i \in \mathbb{R}^{V\times 1}$ is further downsampled through averaging within one ROI to formulate $\overline{\psi}_i \in \mathbb{R}^{N\times1}$, where $V$ represents the number of vertices in the mesh and $N$ denotes the number of ROIs in a brain parcellation. First $J$ downsampled harmonics $\overline{\psi}_J \in \mathbb{R}^{N \times J}$ are selected for learning positional embedding. We incorporate a learnable linear layer to transform geometric harmonics $\overline{\psi}_J$ into positional embeddings $E \in \mathbb{R}^{N \times d}$, where $d$ denotes the embedding dimension of the transformer.

By explicitly encoding the geometric constraints into fMRI representations, we pre-align functional brain organization with cortical structure, enabling more effective integration during subsequent modality fusion. Moreover, embedding this physics-informed inductive bias, derived from population-level observations, can further enhance cross-subject and cross-dataset alignment.

\subsubsection{Temporal Adaptive Patch Embedding (TAPE)}

\begin{wrapfigure}{r}{0.5\textwidth} 
  \centering
  \includegraphics[width=0.5\textwidth]{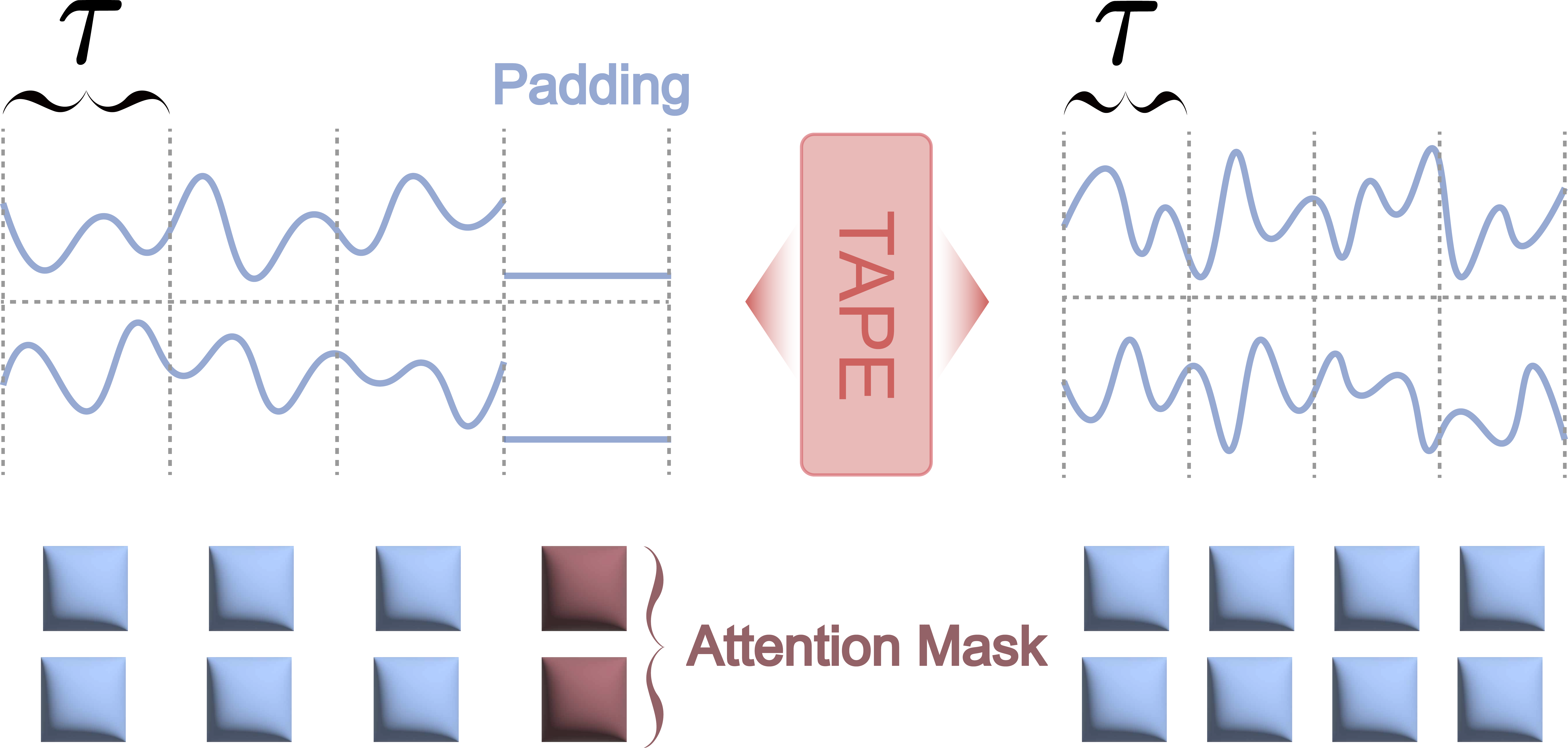} 
  \caption{\textbf{Temporal Adaptive Patch Embedding (TAPE).} Tokens represent the same temporal duration $\tau$. Embedding weights are correspondingly resized. Shorter time series with fewer tokens are zero-padded, with attention masks excluding padded tokens afterwards.}
  \label{fig4}
\end{wrapfigure}

FMRI data collection spans scanners, sites and pulse-sequences that sample the  BOLD signals anywhere from sub-second to several-second TRs. However, existing brain dynamics foundation models rely on a fixed TR for both pretraining and downstream tasks \cite{carobrainlm}, or downsample datasets with higher temporal resolutions to match lower ones \cite{dong2024brain}. This limitation restricts their ability to incorporate diverse datasets during pretraining and adapt flexibly to downstream tasks with varying TRs. Furthermore, downsampling inevitably sacrifices finer temporal details, reducing the richness of information encoded at higher temporal resolutions. The inability of existing models to accommodate heterogeneous TRs stems from the use of a uniform patch size and a patch embedding layer with a fixed size in the transformer. When training on fMRI data with diverse TRs, employing a single patch size leads each patch to inadvertently represent different temporal durations across data. The embedding layer, however, fails to accurately interpret the varying temporal extents of the input. As a result, the tokens passed to the transformer lack consistent and well-defined temporal semantics, introducing ambiguity that hampers effective modeling of brain dynamics. This limitation ultimately degrades the quality of learned representations and impairs downstream performance, particularly in capturing brain–behavior associations.

To overcome this critical limitation, in BrainHarmonix-F, we propose Temporal Adaptive Patch Embedding (TAPE) that dynamically accommodates varying TRs across fMRI data (Figure \ref{fig4}). We first define a consistent temporal duration $\tau$ for any token and a base embedding weight $\omega^* \in \mathbb{R}^{k^*}$ corresponding to the patch size $k^*$. Given an arbitrary fMRI time series with repetition time $\text{TR}=s$, the corresponding patch size $k$ and resized embedding weights $\omega \in \mathbb{R}^{k}$ are computed as:
\vspace{-0.001cm}
\begin{equation}
  k = \operatorname{round} \ \!\Bigl(\frac{\tau}{s}\Bigr), \ \omega=((B^{k^*}_{k})^T)^{\dagger} \cdot \omega^*  
\end{equation}
\vspace{-0.001cm}
where $\omega$ is obtained by pseudoinverse resize (PI-resize) \cite{beyer2023flexivit}, with the linear transformation matrix $B^{k^*}_{k} \in \mathbb{R}^{{k} \times k^*}$. Since different fMRI scans may vary in total duration, patchifying with temporally consistent tokens could result in a varying number of tokens across time series. If the maximum number of tokens per time series across the dataset is $m$, any time series producing fewer tokens ($n < m$) will be zero-padded followed by an attention mask, ensuring tokens derived from padding are excluded from attention computation (Figure \ref{fig4}).

Previously, no established data augmentation techniques existed for fMRI time series; our TAPE uniquely supports arbitrary TRs, enabling the first-ever augmentation by downsampling high-resolution scans into diverse TRs, enhancing model performance (details in Section \ref{4.1}).

\subsection{Multimodal Fusion (MF)}

At BrainHarmonix’s MF stage, we introduce learnable 1D brain hub tokens that serve as a representational bottleneck. These tokens are trained through the attention-based model \cite{vaswani2017attention} to reconstruct both structural and functional latents, effectively capturing the shared information between modalities (Figure \ref{fig2}.B.). 1D brain hub tokens foster a unified and compact latent space that encapsulates the holistic nature of brain morphology and dynamics.

Let $\mathbf{Z}_S \in \mathbb{R}^{N_S \times d}$, $\mathbf{Z}_F \in \mathbb{R}^{N_F \times d}$ be the modalitity-specific latents of one paired T1-fMRI produced by BrainHarmonix-S and BrainHarmonix-F, respectively, where $d$ is the common embedding dimension and $N_S$, $N_F$ are the numbers of tokens in each modality. We introduce a set of $N_H$ learnable continuous-valued 1D brain hub tokens $\mathbf{H}_0 \in \mathbb{R}^{N_H \times d}$, shared by all pairs and optimized jointly with the network in MF. At every forward pass we concatenate the hubs with the two modality sequences and feed the resulting stream to the Harmonizer transformer ($\mathcal{H}$):
\vspace{-0.001cm}
\begin{equation}
    \mathbf{Z}_0 = [\mathbf{H}_0;\mathbf{Z}_S;\mathbf{Z}_F] \in \mathbb{R}^{(N_H+N_S+N_F) \times d}, \ \widetilde{\mathbf{H}}=\mathcal{H}(\mathbf{Z}_0)_{1:N_H} \in \mathbb{R}^{N_H \times d}
\end{equation}
\vspace{-0.001cm}
where $\mathbf{Z}_0$ is the concatenated input to $\mathcal{H}$ and $\widetilde{\mathbf{H}}$ is the hub tokens updated by $\mathcal{H}$. Self-attention within $\mathcal{H}$ allows the 1D tokens to gather information from \emph{both} structural and functional tokens, while also enabling cross-modal interactions between $\mathbf{Z}_S$ and $\mathbf{Z}_F$.

Two lightweight decoders ($\mathcal{D}_S, \mathcal{D}_F$) project $\widetilde{\mathbf{H}}$ back into each modality's latent space (Figure \ref{fig2}.B.). Formally, our training in MF is defined as:
\vspace{-0.001cm}
\begin{equation}
    \min_{\theta_{\mathcal{H}},\;\theta_{\mathcal{D}_S},\;\theta_{\mathcal{D}_F}} \mathcal{L}_{\text{fusion}} = \lVert \mathcal{D}_S(\widetilde{\mathbf{H}}) - \mathbf{Z}_S\rVert_{2}^{2} + \lVert \mathcal{D}_F(\widetilde{\mathbf{H}}) - \mathbf{Z}_F\rVert_{2}^{2}
\end{equation}
\vspace{-0.001cm}
where $\theta_{\mathcal{H}},\;\theta_{\mathcal{D}_S},\;\theta_{\mathcal{D}_F}$ represents the parameters in $\mathcal{H}, \mathcal{D}_S, \text{and} \ \mathcal{D}_F$, respectively. $\lVert \cdot \rVert_{2}^{2}$ denotes the Mean Square Error.

\section{Experiments}

\subsection{Datasets}
\label{4.1}

\textbf{Pre-training.} BrainHarmonix was pretrained on two of the largest-scale neuroimaging datasets: UK Biobank (\href{https://www.ukbiobank.ac.uk/}{UKB}) \cite{bycroft2018uk,miller2016multimodal} and Adolescent Brain Cognitive Development (\href{https://abcdstudy.org/}{ABCD}) \cite{casey2018adolescent}. From UKB, we curated neuroimaging data of 43,112 participants aged between 44 and 83 years, comprising 46,455 T1-weighted MRI scans and 40,162 resting-state fMRI time series (TR = 0.735 s). From ABCD, we included 11,221 participants (aged 8 to 11 years at baseline visit), consisting of 18,139 T1-weighted images and 30,771 resting-state fMRI time series (TR = 0.8 s).

During the UE stage, a total of 64,594 T1-weighted images from both datasets were utilized for BrainHarmonix-S pretraining. For fMRI data augmentation, UKB data underwent temporal downsampling by factors of 1 to 3, resulting in TRs of 0.735 s, 1.47 s, 2.205 s, and 2.94 s. ABCD data were downsampled by factors of 1 to 2, yielding TRs of 0.8 s, 1.6 s, and 2.4 s. Consequently, the total number of pretraining samples for BrainHarmonix-F was 252,961 (UKB: 40,162 × 4; ABCD: 30,771 × 3). In the MF stage, we extracted 69,360 matched T1-fMRI pairs from both datasets (one T1-weighted image could correspond to multiple fMRI runs within a single session). All fMRI data was parcellated into $N=400$ ROIs with Schaefer-400 \cite{schaefer2018local}.
Further details regarding data preprocessing are provided in Appendix \ref{A}.

\textbf{Downstream fine-tuning.} We evaluated BrainHarmonix on six neuroimaging benchmark datasets. Three multi-site datasets focused on neurodevelopmental disorder diagnosis (TR distributions are detailed in Figure \ref{ap1}): Autism Brain Imaging Data Exchange datasets (ABIDE-I and ABIDE-II) for distinguishing Autism Spectrum Disorder (ASD) from controls, and the Attention Deficit Hyperactivity Disorder dataset (ADHD-200) for classifying ADHD versus controls. On the other hand, three datasets assessed neurodegenerative disorders and cognitive function: Parkinson’s Progression Markers Initiative (PPMI) (TR = 2.5s) for four-class classification involving controls, scans without evidence of dopaminergic deficit (SWEDD), prodromal cases, and Parkinson’s disease (PD); Alzheimer's Disease Neuroimaging Initiative (ADNI) (TR = 3.0s) for classification between controls and mild cognitive impairment (MCI); and the Lifespan Human Connectome Project Aging (HCP-A) dataset (TR = 0.8s) for predicting executive function (Flanker task scores). The results were averaged across three independent runs with distinct data splits (train:validation:test = 6:2:2). We adopted the data stratification approach in \cite{kan2022brain} for splitting the neurodevelopmental datasets. Detailed information regarding class distributions and preprocessing procedures for each benchmark dataset can be found in Appendix \ref{A}.

\subsection{Implementation details}
\label{imp}

In UE, we adopted Vision Transformer‑Base (ViT‑B) \cite{dosovitskiy2020image} as backbone for BrainHarmonix-S and BrainHarmonix-F ($\varepsilon_{\scriptscriptstyle S}$ and $\varepsilon_{\scriptscriptstyle F}$). We employed MAE \cite{he2022masked} as the pretraining framework for BrainHarmonix-S with Brain‑JEPA \cite{dong2024brain} for BrainHarmonix-F. In BrainHarmonix-F, it patchified fMRI time series into 1D patches, with the length $k$ dynamically determined by TR. Geometric harmonics and brain gradients \cite{dong2024brain} were each linearly projected, then averaged to produce the final positioning. T1 images were randomly masked while we followed \cite{dong2024brain} to use spatiotemporal masking for fMRI.


In MF, harmonizer ($\mathcal{H}$) was employed with ViT‑B encoder, paired with an MAE‑style decoder ($\mathcal{D}_S \ \& \ \mathcal{D}_F$) whose design matches the encoder’s size \cite{he2022masked}. Throughout both MF and downstream fine-tuning, both BrainHarmonix-S and BrainHarmonix-F were \emph{frozen}, providing modality latents only. For downstream fine-tuning, we average-pooled the brain hub tokens to generate a global multimodal representation followed by a linear projection head. The main results in Section \ref{4.3} were all based on $N_H=128$ 1D tokens. We employed FlashAttention \cite{dao2022flashattention,dao2023flashattention2} in our self-attention implementation to improve computational efficiency and reduce memory usage. Each pre-training process utilized 8 NVIDIA H100 GPUs (80GB). The pretraining of $\mathcal{H}$ with 128 1D tokens took around 10 hours. The readers are refered to Appendix \ref{B} for detailed optimization settings. 

\begin{table}[t!]
  \centering
    \caption{Comparison on neurodevelopmental disorder diagnosis. Results are averages over three random splits (standard deviations in Table \ref{Ap_tab1}). The best results are highlighted in bold (* indicates statistical significance, $p$ < 0.05), and second-best results are underlined. Task details in Section \ref{4.1}.}
  \small
  \setlength{\tabcolsep}{2.8pt}
  \renewcommand{\arraystretch}{1.1}
  \begin{tabularx}{\textwidth}{@{}X c c c@{\hspace{4pt}}| cc| cc| cc@{}}
    \toprule
    \multirow[b]{2}{*}{\textbf{Model}} &
      \multirow[b]{2}{*}{\textbf{Morphology}} &
      \multirow[b]{2}{*}{\textbf{Dynamics}} &
      \multirow[b]{2}{*}{\textbf{Multi-TR}} &
      \multicolumn{2}{c|}{\textbf{ABIDE-I}} &
      \multicolumn{2}{c|}{\textbf{ABIDE-II}} &
      \multicolumn{2}{c}{\textbf{ADHD-200}} \\
    \cmidrule(lr){5-6} \cmidrule(lr){7-8} \cmidrule(lr){9-10}
    & & & & \textbf{ACC}\% & \textbf{F1}\% & \textbf{ACC}\% & \textbf{F1}\% & \textbf{ACC}\% & \textbf{F1}\% \\
    \midrule
    \multicolumn{4}{@{}l}{\emph{Structure-based models}} & \multicolumn{6}{@{}l@{}}{} \\
    \midrule
    BrainMVP\textsuperscript{\scriptsize1} \cite{rui2024brainmvp}   & \cmark & \xmark & N.A.    & 56.50 & 62.46 & 55.71 & 62.16 & \underline{67.72} & 43.97 \\
    BrainMVP\textsuperscript{\scriptsize2} \cite{rui2024brainmvp}     & \cmark & \xmark & N.A.    & 55.06 & 64.43 & 55.63 & 58.76 & 62.59 & 49.95 \\
    \textbf{BrainHarmonix-S}  & \cmark & \xmark & N.A.    & 56.29 & 62.06 & 60.00 & 68.55 & 64.96 & 53.53 \\
    \midrule
    \multicolumn{4}{@{}l}{\emph{Function-based models}} & \multicolumn{6}{@{}l@{}}{} \\
    \midrule
    BrainNetCNN \cite{kawahara2017brainnetcnn}           & \xmark & \xmark & \cmark & 60.49 & 67.13 & 59.71 & 67.27 & 60.54 & 58.62 \\
    BrainGNN \cite{li2021braingnn}              & \xmark & \xmark & \cmark & 56.72 & 65.71 & 58.71 & 66.48 & 62.24 & 60.67 \\
    BrainNetTF \cite{kan2022brain}                   & \xmark & \xmark & \cmark & 56.73 & 64.75 & 62.03 & 67.64 & 61.91 & 62.68 \\
    BrainMass \cite{yang2024brainmass}              & \xmark & \xmark & \cmark & \textbf{65.64*} & 69.07 & 59.35 & 71.86 & 65.99 & 61.27 \\
    BrainLM \cite{carobrainlm}               & \xmark & \cmark & \xmark & -- & -- & -- & -- & -- & -- \\
    Brain-JEPA \cite{dong2024brain}            & \xmark & \cmark & \xmark & -- & -- & -- & -- & -- & -- \\
    \textbf{BrainHarmonix-F}   & \xmark & \cmark & \cmark & 57.39 & \underline{71.24} & \underline{62.90} & \underline{72.76} & 67.69 & \textbf{68.75} \\
    \midrule
    \multicolumn{4}{@{}l}{\emph{Multimodal model}} & \multicolumn{6}{@{}l@{}}{} \\
    \midrule
    \textbf{BrainHarmonix}        & \cmark & \cmark & \cmark & \underline{63.13} & \textbf{72.63*} & \textbf{66.67*} & \textbf{74.88*} & \textbf{70.09*} & \underline{66.72} \\
    \bottomrule
  \end{tabularx}
  \begin{tablenotes}
   \item[*] \hspace{-.5cm}\small \textsuperscript{\scriptsize1}UniFormer \cite{li2023uniformer} as backbone; \textsuperscript{\scriptsize2}UNET3D \cite{ronneberger2015u} as backbone.
    \end{tablenotes}
  \label{tab1}
\end{table}

\subsection{Main results}
\label{4.3}
BrainHarmonix demonstrated strong generalization capabilities across neurodevelopmental and neurodegenerative disorder diagnoses as well as cognition prediction (Table \ref{tab1}, \ref{tab2}). As the first multimodal brain foundation model, BrainHarmonix was benchmarked against both structure-based and function-based neuroimaging models. For structure-based comparisons, we included BrainMVP \cite{rui2024brainmvp}, a state-of-the-art structural foundation model originally designed for multi-parametric MRI. Given its incompatibility of pretraining with T1 images only, we adopted BrainMVP's pretrained weights and fine-tune it on downstream datasets. Many task-specific models for fMRI based on deep learning were proposed before foundation models emerged. These models could only be applied to specific tasks rather than a wide range of downstream applications \cite{kawahara2017brainnetcnn, li2021braingnn, kan2022brain, dong2023beyond, dong2022coop}. For functional comparisons, BrainHarmonix was evaluated against both task-specific (BrainNetCNN \cite{kawahara2017brainnetcnn}, BrainGNN \cite{li2021braingnn}, and BrainNetTF \cite{kan2022brain}) and foundational fMRI models (BrainMass \cite{yang2024brainmass}, BrainLM \cite{carobrainlm}, and Brain-JEPA \cite{dong2024brain}). Previous brain dynamics foundation models, including BrainLM and Brain-JEPA, are not able to handle heterogeneous TRs. Consequently, these models were only assessed on datasets with homogeneous TRs (PPMI, ADNI, and HCP-A) following their original downsampling strategies (Table \ref{tab2}). BrainMass was pretrained on our pretraining datasets following the original settings.

Overall, BrainHarmonix consistently outperformed both structure-based and function-based models. Among its ablations, BrainHarmonix-F, which exclusively captures functional dynamics, achieved superior performance compared to existing fMRI models, highlighting the effectiveness of modeling heterogeneous dynamics from large-scale neuroimaging data. On the other hand, BrainHarmonix-S achieved performance that is superior or comparable to BrainMVP through an MAE framework, pretrained on large-scale T1 datasets without multi-parametric MRI. The performance of BrainHarmonix-S can be attributed to its pretraining on a significantly larger T1 dataset, resulting in more robust brain morphological representations. The further improvements observed after structural-functional fusion in BrainHarmonix underscore the significance of integrating multimodal heuristics for comprehensive human brain representation.

\begin{table}[t!]
  \centering
    \caption{Comparison on neurodegenerative disease diagnosis and cognition prediction (standard deviations in Table \ref{Ap_tab2}). Task details in Section \ref{4.1}.}
  \small
  \setlength{\tabcolsep}{3pt}
  \renewcommand{\arraystretch}{1.1}
  \begin{tabularx}{\textwidth}{@{}X c c c@{\hspace{4pt}}| cc| cc| cc@{}}
    \toprule
    \multirow[b]{2}{*}{\textbf{Model}} &
      \multirow[b]{2}{*}{\textbf{Morphology}} &
      \multirow[b]{2}{*}{\textbf{Dynamics}} &
      \multirow[b]{2}{*}{\textbf{Multi-TR}} &
      \multicolumn{2}{c|}{\textbf{PPMI}} &
      \multicolumn{2}{c|}{\textbf{ADNI}} &
      \multicolumn{2}{c}{\textbf{HCP-A}} \\
    \cmidrule(lr){5-6} \cmidrule(lr){7-8} \cmidrule(lr){9-10}
    & & & & \textbf{ACC}\% & \textbf{F1}\% & \textbf{ACC}\% & \textbf{F1}\% & \textbf{MAE} & $\rho$ \\
    \midrule
    \multicolumn{4}{@{}l}{\emph{Structure-based models}} & \multicolumn{6}{@{}l@{}}{} \\
    \midrule
    BrainMVP\textsuperscript{\scriptsize1} \cite{rui2024brainmvp}  & \cmark & \xmark & N.A.    & 58.94 & 50.71 & 57.41 & 54.88 & 5.80 & 0.25 \\
    BrainMVP\textsuperscript{\scriptsize2} \cite{rui2024brainmvp}     & \cmark & \xmark & N.A.    & 55.04 & 40.82 & 60.61 & 44.67 & \textbf{5.39*} & 0.36 \\
    \textbf{BrainHarmonix-S}  & \cmark & \xmark & N.A.    & 59.69 & 51.04 & 57.59 & 56.09 & 6.05 & \underline{0.38} \\
    \midrule
    \multicolumn{4}{@{}l}{\emph{Function-based models}} & \multicolumn{6}{@{}l@{}}{} \\
    \midrule
    BrainNetCNN \cite{kawahara2017brainnetcnn}           & \xmark & \xmark & \cmark & 56.59 & 46.59 & 56.57 & 54.59 & 6.82 & 0.23 \\
    BrainGNN \cite{li2021braingnn}              & \xmark & \xmark & \cmark & 58.14 & 47.79 & 58.59 & 57.27 & 6.78 & 0.22 \\
    BrainNetTF \cite{kan2022brain}                   & \xmark & \xmark & \cmark & 58.92 & 48.56 & 60.61 & 58.00 & 6.70 & 0.25 \\
    BrainMass \cite{yang2024brainmass}             & \xmark & \xmark & \cmark & 59.77 & 48.31 & 59.60 & 56.73 & 6.45 & 0.28 \\
    BrainLM \cite{carobrainlm}               & \xmark & \cmark & \xmark & 53.49 & 44.58 & 57.58 & 59.57 & 7.03 & 0.25 \\
    Brain-JEPA \cite{dong2024brain}            & \xmark & \cmark & \xmark & 60.36 & 48.76 & 59.60 & 60.78 & \underline{5.62} & 0.26 \\
    \textbf{BrainHarmonix-F}  & \xmark & \cmark & \cmark & \underline{62.79} & \underline{52.90} & \underline{61.62} & \underline{64.80} & 5.77 & 0.30 \\
    \midrule
    \multicolumn{4}{@{}l}{\emph{Multimodal model}} & \multicolumn{6}{@{}l@{}}{} \\
    \midrule
    \textbf{BrainHarmonix}        & \cmark & \cmark & \cmark & \textbf{64.34*} & \textbf{56.40*} & \textbf{64.65*} & \textbf{68.75*} & 6.56 & \textbf{0.42*} \\
    \bottomrule
  \end{tabularx}
  \begin{tablenotes}
   \item[*] \hspace{-.5cm}\small \textsuperscript{\scriptsize1}UniFormer \cite{li2023uniformer} as backbone; \textsuperscript{\scriptsize2}UNET3D \cite{ronneberger2015u} as backbone.
    \end{tablenotes}
  \label{tab2}
\end{table}

\begin{figure}[!t]
    \centering
    \includegraphics[width=\columnwidth]{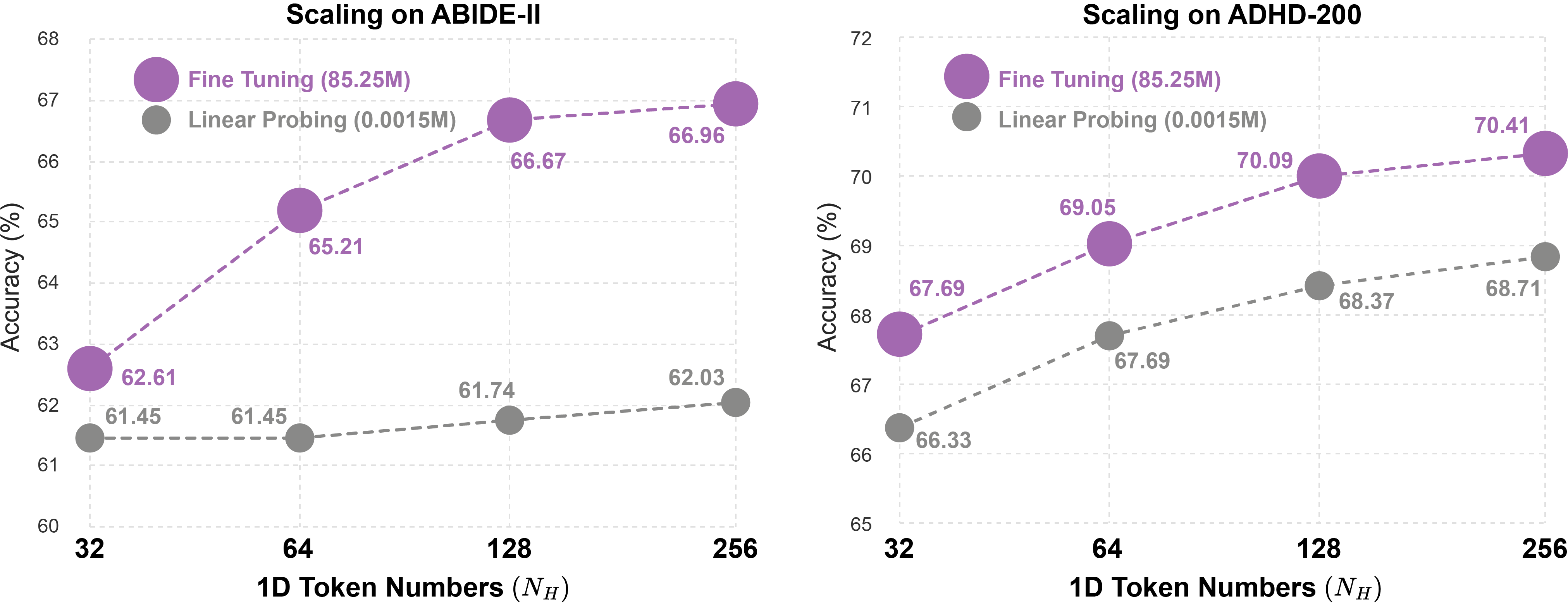}
    \caption{\textbf{Scaling across different numbers of 1D tokens in fine-tuning and linear probing.} We plot both fine-tuning (85.25M trainable parameters) and linear probing (0.0015M trainable parameters). Increasing the token count from 32 to 256 steadily improves accuracy before reaching a plateau. Notably, \textbf{\textcolor{RoyalBlue}{\emph{even our linear-probing approach achieves very strong performance, surpassing prior state-of-the-art results}}} despite using a minimal set of learnable parameters.}
    \label{scaling_fig}
\end{figure}

\begin{figure}[t!]
    \centering
    \includegraphics[width=\columnwidth]{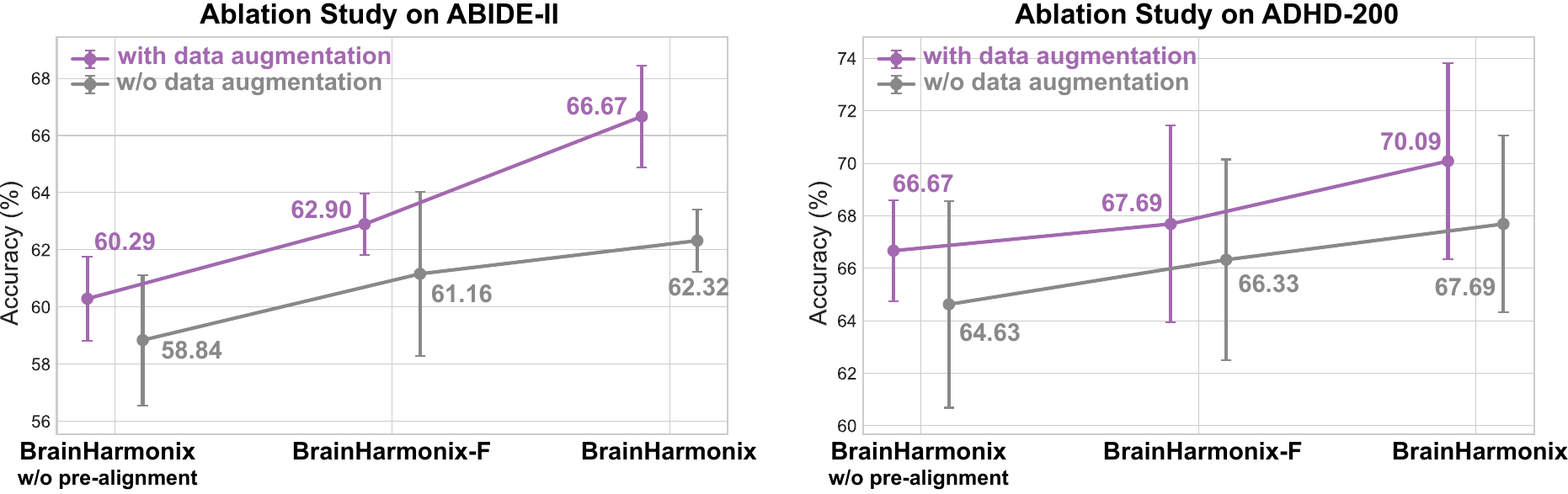}
    \caption{\textbf{Ablation study showing the impact of pre-alignment, data augmentation, and multimodal fusion.} In BrainHarmonix-F without pre-alignment, we sticked to the brain gradient positioning in Brain-JEPA, while the others (BrainHarmonix-F and BrainHarmonix) were injected with geometric harmonics on top of brain gradient (Section \ref{imp}). Results were averaged over three random splits with error bars indicating standard deviations.}
    \label{ablation}
\end{figure}

\subsection{1D token scaling \& linear probing}
\label{4.4}

 We investigated how performance changed when we varied the number of 1D tokens from 32 to 256 and evaluated both fine-tuning and linear probing (Figure \ref{scaling_fig}). As the number of tokens grows, model accuracy consistently increases - highlighting the benefit of richer token-based representations - yet the gains begin to saturate from 128 to 256 tokens. Notably, our linear-probing approach, which uses only a simple linear head on top of the frozen BrainHarmonix, already achieves performance that ourperforms previous advanced baselines. This underscores the strength of the learned brain representations and their capacity to generalize even with minimal downstream adaptation.

\subsection{Ablation study}
\label{4.5}

We compared BrainHarmonix with its ablated versions in Figure \ref{ablation}. The comparison between the purple ("with data augmentation") and gray ("w/o data augmentation") lines demonstrates the consistent performance gains from augmenting fMRI time series through multi-TR downsampling. It illustrates the effectiveness of enriching the model’s temporal representation. Comparing the center bars ("BrainHarmonix-F") to the left ("BrainHarmonix-F w/o pre-alignment") further reveals that pre-alignment of fMRI signals to cortical geometry significantly boosts performance. Finally, the rightmost bar (“BrainHarmonix”) underscores the value of fusing structural and functional information: integrating these two complementary views of the brain yields the highest accuracy.

\subsection{Latent space analyses and interpretation}
\label{4.6}

\begin{table}[h]
\centering
\caption{Comparison of significant modes in t-SNE dimensions}
\begin{tabular}{llccc}
\toprule
\textbf{Dimension} & \textbf{Model} & \textbf{\# Significant Modes} & \textbf{Avg. P-value} & \textbf{Avg. Correlation} \\
\midrule
\multicolumn{5}{l}{\textit{Dim1 in t-SNE}} \\
\midrule
& Brain-JEPA \cite{dong2024brain} & 7 & 0.00769 & 0.1562 \\
& BrainHarmonix-F & \textbf{12} & \textbf{0.00456} & \textbf{0.1717} \\
\midrule
\multicolumn{5}{l}{\textit{Dim2 in t-SNE}} \\
\midrule
& Brain-JEPA \cite{dong2024brain} & 8 & 0.0115 & 0.1506 \\
& BrainHarmonix-F & \textbf{15} & \textbf{0.00477} & \textbf{0.1726} \\
\bottomrule
\end{tabular}
\begin{tablenotes}
   \item[*] \hspace{-.3cm}\small \# = number of; Avg. = average; Dim = dimension.
    \end{tablenotes}
\label{latent}
\end{table}

We extracted fMRI embeddings from BrainHarmonix-F and Brain-JEPA (400 ROIs, each represented by a 768-dimensional embedding) and applied t-SNE to project these embeddings onto a 2D plane. Specifically, we correlated each dimension of the t-SNE embedding with each of the 200 geometric harmonic modes across 400 ROIs. Compared to Brain-JEPA, our geometry-constrained embeddings exhibit a greater number of significantly correlated modes (p<0.05), with higher correlation strengths and significance levels on the top 5 most significant modes (Table \ref{latent}). On the other hand, we applied the Fisher r-to-z transformation to all correlations from the 200 harmonics for each model and conducted a two-sample t-test. Results demonstrate that correlations from our model are significantly higher overall. The correlation strength and statistical significance from the top modes, along with the overall comparison, confirm that our model is constrained by structural information more than Brain-JEPA. 

We examined the attention patterns between the 128 learned 1D tokens and the modality-specific tokens (400 fMRI ROI + 1200 T1 tokens) in ASD diagnosis using ABIDE-II data. For the 400 fMRI ROI tokens, each is obtained by averaging all tokens within the corresponding ROI. We found differentiation in modality attention among the 1D tokens: 93/128 tokens attended exclusively to fMRI, 30/128 exclusively to T1, and 5/128 tokens exhibited cross-modal attention. For cross-modal tokens, we found that they exhibited key structure-function coupling such as medial prefrontal cortex in brain morphometry and default model network in brain dynamics, which have previously been demonstrated in the literature to be associated with ASD.

For the 93 fMRI-specific tokens, further analysis revealed network-level functional differentiation relevant to ASD behavioral traits. Specifically, 60/93 were network-specific, predominantly focusing on a single brain network (with >70\% salient ROIs within one network), while the remaining 33 were identified as “bridge” tokens capturing interactions across multiple networks. Among the most salient network-specific tokens, temporoparietal network (implicated in social perception and language processing deficits), somatomotor network (associated with sensorimotor integration impairments), and default mode network (linked to mentalizing deficits) emerged prominently. The identified “bridge” tokens primarily captured interactions involving default, limbic, and control networks, reflecting impaired integration across sensorimotor, socioemotional, and higher-order cognitive processes - a mechanism implicated in pathophysiology of ASD. 

\section{Conclusion}

In this paper, we introduced \textbf{Brain Harmony (BrainHarmonix)}, the first multimodal brain foundation model that unifies structural morphology and functional dynamics into compact 1D tokens. By integrating geometric harmonics for structural-functional pre-alignment and introducing the Temporal Adaptive Patch Embedding (TAPE) for handling heterogeneous repetition times (TRs) in fMRI datasets, BrainHarmonix effectively bridges critical gaps existing in previous brain representation learning frameworks. Our approach provides a unified, expressive latent space that significantly enhances the representation of complex brain morphology and dynamics. Extensive experiments demonstrated BrainHarmonix's superior generalization across diverse neuroimaging benchmarks, consistently outperforming state-of-the-art models in neurodevelopmental and neurodegenerative disorder classification and cognition prediction. BrainHarmonix is positioned to fundamentally advance AI-driven neuroscience research and clinical applications through multimodal neuroimaging.

\section{Limitations and Future Work}
\label{lim}

We acknowledge several limitations in our study and highlight directions for future work. First, although BrainHarmonix was pretrained on the largest curation of structural-functional neuroimaging datasets to date, the age distribution of the data could be further expanded to better represent the entire human lifespan, particularly infancy and young adulthood. On the other hand,  jointly optimizing the unimodal encoders and the fusion module could potentially lead to further performance gains. Exploring efficient training strategies for such long-sequence transformer models, particularly in the context of neuroimaging, is a promising direction. Beyond the demonstrated gains in neuropsychiatric disease diagnosis and cognition prediction, our multimodal brain foundation model, BrainHarmonix, holds promise -if further developed- as an AI-driven brain digital twin: a neuroscientific tool capable of validating and potentially uncovering  novel neuroscience insights such as specific brain structure-function coupling related to human behavioral phenotypes. Realizing this potential, however, will require rigorous evaluation across diverse tasks and populations as well as systematic investigation into the model's interpretability and translational pathways.

\newpage

\section*{Acknowledgements}

This study was supported by the Singapore National Medical Research Council (NMRC/OFLCG19May-0035, NMRC/CIRG/1485/2018, NMRC/CSA-SI/0007/2016, NMRC/MOH-00707-01, NMRC/CG/435 M009/2017-NUH/NUHS, CIRG21nov-0007, HLCA23Feb-0004, and OFIRG24Jul-0049), RIE2020 AME Programmatic Fund from A*STAR, Singapore (No. A20G8b0102), Ministry of Education (MOE-T2EP40120-0007 \& T2EP2-0223-0025, MOE-T2EP20220-0001), and Yong Loo Lin School of Medicine Research Core Funding, National University of Singapore, Singapore. We would like to acknowledge that GPU resource used in this work was supported by NUS IT’s Research Computing group using grant numbers
NUSREC-HPC-00001.



\bibliographystyle{unsrtnat}
\bibliography{ref}

\begin{thebibliography}{46}
\providecommand{\natexlab}[1]{#1}
\providecommand{\url}[1]{\texttt{#1}}
\expandafter\ifx\csname urlstyle\endcsname\relax
  \providecommand{\doi}[1]{doi: #1}\else
  \providecommand{\doi}{doi: \begingroup \urlstyle{rm}\Url}\fi

\bibitem[Logothetis(2008)]{logothetis2008we}
Nikos~K Logothetis.
\newblock What we can do and what we cannot do with fmri.
\newblock \emph{Nature}, 453\penalty0 (7197):\penalty0 869--878, 2008.

\bibitem[Poldrack and Farah(2015)]{poldrack2015progress}
Russell~A Poldrack and Martha~J Farah.
\newblock Progress and challenges in probing the human brain.
\newblock \emph{Nature}, 526\penalty0 (7573):\penalty0 371--379, 2015.

\bibitem[Calhoun and Sui(2016)]{calhoun2016multimodal}
Vince~D Calhoun and Jing Sui.
\newblock Multimodal fusion of brain imaging data: a key to finding the missing link (s) in complex mental illness.
\newblock \emph{Biological psychiatry: cognitive neuroscience and neuroimaging}, 1\penalty0 (3):\penalty0 230--244, 2016.

\bibitem[Caro et~al.()Caro, de~Oliveira~Fonseca, Rizvi, Rosati, Averill, Cross, Mittal, Zappala, Dhodapkar, Abdallah, et~al.]{carobrainlm}
Josue~Ortega Caro, Antonio~Henrique de~Oliveira~Fonseca, Syed~A Rizvi, Matteo Rosati, Christopher Averill, James~L Cross, Prateek Mittal, Emanuele Zappala, Rahul~Madhav Dhodapkar, Chadi Abdallah, et~al.
\newblock Brainlm: A foundation model for brain activity recordings.
\newblock In \emph{The Twelfth International Conference on Learning Representations}.

\bibitem[Dong et~al.(2024{\natexlab{a}})Dong, Li, Wu, Nguyen, Chong, Ji, Tong, Chen, and Zhou]{dong2024brain}
Zijian Dong, Ruilin Li, Yilei Wu, Thuan~Tinh Nguyen, Joanna Chong, Fang Ji, Nathanael Tong, Christopher Chen, and Juan~Helen Zhou.
\newblock Brain-jepa: Brain dynamics foundation model with gradient positioning and spatiotemporal masking.
\newblock \emph{Advances in Neural Information Processing Systems}, 37:\penalty0 86048--86073, 2024{\natexlab{a}}.

\bibitem[Rui et~al.(2024)Rui, Chen, Tang, Wang, Liu, Zhang, and Wang]{rui2024brainmvp}
Shaohao Rui, Lingzhi Chen, Zhenyu Tang, Lilong Wang, Mianxin Liu, Shaoting Zhang, and Xiaosong Wang.
\newblock Brainmvp: Multi-modal vision pre-training for brain image analysis using multi-parametric mri.
\newblock \emph{arXiv preprint arXiv:2410.10604}, 2024.

\bibitem[Yang et~al.(2024)Yang, Ye, Su, Zhang, Chang, Chen, Chan, Yu, and Ma]{yang2024brainmass}
Yanwu Yang, Chenfei Ye, Guinan Su, Ziyao Zhang, Zhikai Chang, Hairui Chen, Piu Chan, Yue Yu, and Ting Ma.
\newblock Brainmass: Advancing brain network analysis for diagnosis with large-scale self-supervised learning.
\newblock \emph{IEEE Transactions on Medical Imaging}, 2024.

\bibitem[Pang et~al.(2023)Pang, Aquino, Oldehinkel, Robinson, Fulcher, Breakspear, and Fornito]{pang2023geometric}
James~C Pang, Kevin~M Aquino, Marianne Oldehinkel, Peter~A Robinson, Ben~D Fulcher, Michael Breakspear, and Alex Fornito.
\newblock Geometric constraints on human brain function.
\newblock \emph{Nature}, 618\penalty0 (7965):\penalty0 566--574, 2023.

\bibitem[consortium(2012)]{adhd2012adhd}
ADHD-200 consortium.
\newblock The adhd-200 consortium: a model to advance the translational potential of neuroimaging in clinical neuroscience.
\newblock \emph{Frontiers in systems neuroscience}, 6:\penalty0 62, 2012.

\bibitem[Di~Martino et~al.(2014)Di~Martino, Yan, Li, Denio, Castellanos, Alaerts, Anderson, Assaf, Bookheimer, Dapretto, et~al.]{di2014autism}
Adriana Di~Martino, Chao-Gan Yan, Qingyang Li, Erin Denio, Francisco~X Castellanos, Kaat Alaerts, Jeffrey~S Anderson, Michal Assaf, Susan~Y Bookheimer, Mirella Dapretto, et~al.
\newblock The autism brain imaging data exchange: towards a large-scale evaluation of the intrinsic brain architecture in autism.
\newblock \emph{Molecular psychiatry}, 19\penalty0 (6):\penalty0 659--667, 2014.

\bibitem[Di~Martino et~al.(2017)Di~Martino, O’connor, Chen, Alaerts, Anderson, Assaf, Balsters, Baxter, Beggiato, Bernaerts, et~al.]{di2017enhancing}
Adriana Di~Martino, David O’connor, Bosi Chen, Kaat Alaerts, Jeffrey~S Anderson, Michal Assaf, Joshua~H Balsters, Leslie Baxter, Anita Beggiato, Sylvie Bernaerts, et~al.
\newblock Enhancing studies of the connectome in autism using the autism brain imaging data exchange ii.
\newblock \emph{Scientific data}, 4\penalty0 (1):\penalty0 1--15, 2017.

\bibitem[Chang and Glover(2010)]{chang2010time}
Catie Chang and Gary~H Glover.
\newblock Time--frequency dynamics of resting-state brain connectivity measured with fmri.
\newblock \emph{Neuroimage}, 50\penalty0 (1):\penalty0 81--98, 2010.

\bibitem[Calhoun et~al.(2014)Calhoun, Miller, Pearlson, and Adal{\i}]{calhoun2014chronnectome}
Vince~D Calhoun, Robyn Miller, Godfrey Pearlson, and Tulay Adal{\i}.
\newblock The chronnectome: time-varying connectivity networks as the next frontier in fmri data discovery.
\newblock \emph{Neuron}, 84\penalty0 (2):\penalty0 262--274, 2014.

\bibitem[Park et~al.(2025)Park, Park, Bang, Choi, Chung, Kim, and Lee]{park2025foundational}
Joonhyeong Park, Byoungwoo Park, Chang-Bae Bang, Jungwon Choi, Hyungjin Chung, Byung-Hoon Kim, and Juho Lee.
\newblock A foundational brain dynamics model via stochastic optimal control.
\newblock \emph{arXiv preprint arXiv:2502.04892}, 2025.

\bibitem[He et~al.(2022)He, Chen, Xie, Li, Doll{\'a}r, and Girshick]{he2022masked}
Kaiming He, Xinlei Chen, Saining Xie, Yanghao Li, Piotr Doll{\'a}r, and Ross Girshick.
\newblock Masked autoencoders are scalable vision learners.
\newblock In \emph{Proceedings of the IEEE/CVF conference on computer vision and pattern recognition}, pages 16000--16009, 2022.

\bibitem[Beyer et~al.(2023)Beyer, Izmailov, Kolesnikov, Caron, Kornblith, Zhai, Minderer, Tschannen, Alabdulmohsin, and Pavetic]{beyer2023flexivit}
Lucas Beyer, Pavel Izmailov, Alexander Kolesnikov, Mathilde Caron, Simon Kornblith, Xiaohua Zhai, Matthias Minderer, Michael Tschannen, Ibrahim Alabdulmohsin, and Filip Pavetic.
\newblock Flexivit: One model for all patch sizes.
\newblock In \emph{Proceedings of the IEEE/CVF Conference on Computer Vision and Pattern Recognition}, pages 14496--14506, 2023.

\bibitem[Vaswani et~al.(2017)Vaswani, Shazeer, Parmar, Uszkoreit, Jones, Gomez, Kaiser, and Polosukhin]{vaswani2017attention}
Ashish Vaswani, Noam Shazeer, Niki Parmar, Jakob Uszkoreit, Llion Jones, Aidan~N Gomez, {\L}ukasz Kaiser, and Illia Polosukhin.
\newblock Attention is all you need.
\newblock \emph{Advances in neural information processing systems}, 30, 2017.

\bibitem[Bycroft et~al.(2018)Bycroft, Freeman, Petkova, Band, Elliott, Sharp, Motyer, Vukcevic, Delaneau, O’Connell, et~al.]{bycroft2018uk}
Clare Bycroft, Colin Freeman, Desislava Petkova, Gavin Band, Lloyd~T Elliott, Kevin Sharp, Allan Motyer, Damjan Vukcevic, Olivier Delaneau, Jared O’Connell, et~al.
\newblock The uk biobank resource with deep phenotyping and genomic data.
\newblock \emph{Nature}, 562\penalty0 (7726):\penalty0 203--209, 2018.

\bibitem[Miller et~al.(2016)Miller, Alfaro-Almagro, Bangerter, Thomas, Yacoub, Xu, Bartsch, Jbabdi, Sotiropoulos, Andersson, et~al.]{miller2016multimodal}
Karla~L Miller, Fidel Alfaro-Almagro, Neal~K Bangerter, David~L Thomas, Essa Yacoub, Junqian Xu, Andreas~J Bartsch, Saad Jbabdi, Stamatios~N Sotiropoulos, Jesper~LR Andersson, et~al.
\newblock Multimodal population brain imaging in the uk biobank prospective epidemiological study.
\newblock \emph{Nature neuroscience}, 19\penalty0 (11):\penalty0 1523--1536, 2016.

\bibitem[Casey et~al.(2018)Casey, Cannonier, Conley, Cohen, Barch, Heitzeg, Soules, Teslovich, Dellarco, Garavan, et~al.]{casey2018adolescent}
Betty~Jo Casey, Tariq Cannonier, May~I Conley, Alexandra~O Cohen, Deanna~M Barch, Mary~M Heitzeg, Mary~E Soules, Theresa Teslovich, Danielle~V Dellarco, Hugh Garavan, et~al.
\newblock The adolescent brain cognitive development (abcd) study: imaging acquisition across 21 sites.
\newblock \emph{Developmental cognitive neuroscience}, 32:\penalty0 43--54, 2018.

\bibitem[Schaefer et~al.(2018)Schaefer, Kong, Gordon, Laumann, Zuo, Holmes, Eickhoff, and Yeo]{schaefer2018local}
Alexander Schaefer, Ru~Kong, Evan~M Gordon, Timothy~O Laumann, Xi-Nian Zuo, Avram~J Holmes, Simon~B Eickhoff, and BT~Thomas Yeo.
\newblock Local-global parcellation of the human cerebral cortex from intrinsic functional connectivity mri.
\newblock \emph{Cerebral cortex}, 28\penalty0 (9):\penalty0 3095--3114, 2018.

\bibitem[Kan et~al.(2022)Kan, Dai, Cui, Zhang, Guo, and Yang]{kan2022brain}
Xuan Kan, Wei Dai, Hejie Cui, Zilong Zhang, Ying Guo, and Carl Yang.
\newblock Brain network transformer.
\newblock \emph{Advances in Neural Information Processing Systems}, 35:\penalty0 25586--25599, 2022.

\bibitem[Dosovitskiy et~al.(2020)Dosovitskiy, Beyer, Kolesnikov, Weissenborn, Zhai, Unterthiner, Dehghani, Minderer, Heigold, Gelly, et~al.]{dosovitskiy2020image}
Alexey Dosovitskiy, Lucas Beyer, Alexander Kolesnikov, Dirk Weissenborn, Xiaohua Zhai, Thomas Unterthiner, Mostafa Dehghani, Matthias Minderer, Georg Heigold, Sylvain Gelly, et~al.
\newblock An image is worth 16x16 words: Transformers for image recognition at scale.
\newblock \emph{arXiv preprint arXiv:2010.11929}, 2020.

\bibitem[Dao et~al.(2022)Dao, Fu, Ermon, Rudra, and R{\'e}]{dao2022flashattention}
Tri Dao, Daniel~Y. Fu, Stefano Ermon, Atri Rudra, and Christopher R{\'e}.
\newblock Flash{A}ttention: Fast and memory-efficient exact attention with {IO}-awareness.
\newblock In \emph{Advances in Neural Information Processing Systems (NeurIPS)}, 2022.

\bibitem[Dao(2024)]{dao2023flashattention2}
Tri Dao.
\newblock Flash{A}ttention-2: Faster attention with better parallelism and work partitioning.
\newblock In \emph{International Conference on Learning Representations (ICLR)}, 2024.

\bibitem[Kawahara et~al.(2017)Kawahara, Brown, Miller, Booth, Chau, Grunau, Zwicker, and Hamarneh]{kawahara2017brainnetcnn}
Jeremy Kawahara, Colin~J Brown, Steven~P Miller, Brian~G Booth, Vann Chau, Ruth~E Grunau, Jill~G Zwicker, and Ghassan Hamarneh.
\newblock Brainnetcnn: Convolutional neural networks for brain networks; towards predicting neurodevelopment.
\newblock \emph{NeuroImage}, 146:\penalty0 1038--1049, 2017.

\bibitem[Li et~al.(2021)Li, Zhou, Dvornek, Zhang, Gao, Zhuang, Scheinost, Staib, Ventola, and Duncan]{li2021braingnn}
Xiaoxiao Li, Yuan Zhou, Nicha Dvornek, Muhan Zhang, Siyuan Gao, Juntang Zhuang, Dustin Scheinost, Lawrence~H Staib, Pamela Ventola, and James~S Duncan.
\newblock Braingnn: Interpretable brain graph neural network for fmri analysis.
\newblock \emph{Medical Image Analysis}, 74:\penalty0 102233, 2021.

\bibitem[Li et~al.(2023)Li, Wang, Zhang, Gao, Song, Liu, Li, and Qiao]{li2023uniformer}
Kunchang Li, Yali Wang, Junhao Zhang, Peng Gao, Guanglu Song, Yu~Liu, Hongsheng Li, and Yu~Qiao.
\newblock Uniformer: Unifying convolution and self-attention for visual recognition.
\newblock \emph{IEEE Transactions on Pattern Analysis and Machine Intelligence}, 45\penalty0 (10):\penalty0 12581--12600, 2023.

\bibitem[Ronneberger et~al.(2015)Ronneberger, Fischer, and Brox]{ronneberger2015u}
Olaf Ronneberger, Philipp Fischer, and Thomas Brox.
\newblock U-net: Convolutional networks for biomedical image segmentation.
\newblock In \emph{Medical image computing and computer-assisted intervention--MICCAI 2015: 18th international conference, Munich, Germany, October 5-9, 2015, proceedings, part III 18}, pages 234--241. Springer, 2015.

\bibitem[Dong et~al.(2023)Dong, Wu, Xiao, Chong, Jin, and Zhou]{dong2023beyond}
Zijian Dong, Yilei Wu, Yu~Xiao, Joanna Su~Xian Chong, Yueming Jin, and Juan~Helen Zhou.
\newblock Beyond the snapshot: Brain tokenized graph transformer for longitudinal brain functional connectome embedding.
\newblock In \emph{International Conference on Medical Image Computing and Computer-Assisted Intervention}, pages 348--357. Springer, 2023.

\bibitem[Dong et~al.(2022)Dong, Chong, Kok, and Zhou]{dong2022coop}
Zijian Dong, Joanna Su~Xian Chong, Bing~Cai Kok, and Juan~Helen Zhou.
\newblock Coop-dhgnn: a framework for joint classification and prediction of brain functional connectivity using sparse trajectory dataset with application to early dementia.
\newblock In \emph{2022 IEEE International Conference on Big Data (Big Data)}, pages 4972--4978. IEEE, 2022.

\bibitem[Leonardsen et~al.(2022)Leonardsen, Peng, Kaufmann, Agartz, Andreassen, Celius, Espeseth, Harbo, H{\o}gest{\o}l, De~Lange, et~al.]{leonardsen2022deep}
Esten~H Leonardsen, Han Peng, Tobias Kaufmann, Ingrid Agartz, Ole~A Andreassen, Elisabeth~Gulowsen Celius, Thomas Espeseth, Hanne~F Harbo, Einar~A H{\o}gest{\o}l, Ann-Marie De~Lange, et~al.
\newblock Deep neural networks learn general and clinically relevant representations of the ageing brain.
\newblock \emph{NeuroImage}, 256:\penalty0 119210, 2022.

\bibitem[S{\'e}gonne et~al.(2004)S{\'e}gonne, Dale, Busa, Glessner, Salat, Hahn, and Fischl]{segonne2004hybrid}
Florent S{\'e}gonne, Anders~M Dale, Evelina Busa, Maureen Glessner, David Salat, Horst~Karl Hahn, and Bruce Fischl.
\newblock A hybrid approach to the skull stripping problem in mri.
\newblock \emph{Neuroimage}, 22\penalty0 (3):\penalty0 1060--1075, 2004.

\bibitem[Jenkinson et~al.(2012)Jenkinson, Beckmann, Behrens, Woolrich, and Smith]{jenkinson2012fsl}
Mark Jenkinson, Christian~F Beckmann, Timothy~EJ Behrens, Mark~W Woolrich, and Stephen~M Smith.
\newblock Fsl.
\newblock \emph{Neuroimage}, 62\penalty0 (2):\penalty0 782--790, 2012.

\bibitem[Jenkinson and Smith(2001)]{jenkinson2001global}
Mark Jenkinson and Stephen Smith.
\newblock A global optimisation method for robust affine registration of brain images.
\newblock \emph{Medical image analysis}, 5\penalty0 (2):\penalty0 143--156, 2001.

\bibitem[Di~Biase et~al.(2023)Di~Biase, Smith, Zalesky, Seguin, et~al.]{di2023connectomes}
Maria~A Di~Biase, Robert~E Smith, Andrew Zalesky, Caio Seguin, et~al.
\newblock Connectomes for 40,000 uk biobank participants: A multi-modal, multi-scale brain network resource.
\newblock \emph{NeuroImage}, 283:\penalty0 120407, 2023.

\bibitem[Loshchilov and Hutter(2019)]{loshchilov2018decoupled}
Ilya Loshchilov and Frank Hutter.
\newblock Decoupled weight decay regularization.
\newblock In \emph{International Conference on Learning Representations}, 2019.
\newblock URL \url{https://openreview.net/forum?id=Bkg6RiCqY7}.

\bibitem[Assran et~al.(2023)Assran, Duval, Misra, Bojanowski, Vincent, Rabbat, LeCun, and Ballas]{assran2023self}
Mahmoud Assran, Quentin Duval, Ishan Misra, Piotr Bojanowski, Pascal Vincent, Michael Rabbat, Yann LeCun, and Nicolas Ballas.
\newblock Self-supervised learning from images with a joint-embedding predictive architecture.
\newblock In \emph{Proceedings of the IEEE/CVF Conference on Computer Vision and Pattern Recognition}, pages 15619--15629, 2023.

\bibitem[Marek et~al.(2011)Marek, Jennings, Lasch, Siderowf, Tanner, Simuni, Coffey, Kieburtz, Flagg, Chowdhury, et~al.]{marek2011parkinson}
Kenneth Marek, Danna Jennings, Shirley Lasch, Andrew Siderowf, Caroline Tanner, Tanya Simuni, Chris Coffey, Karl Kieburtz, Emily Flagg, Sohini Chowdhury, et~al.
\newblock The parkinson progression marker initiative (ppmi).
\newblock \emph{Progress in neurobiology}, 95\penalty0 (4):\penalty0 629--635, 2011.

\bibitem[Xu et~al.(2023)Xu, Yang, Huang, Gururajapathy, Ke, Qiao, Wang, Kumar, McGeown, and Kwon]{xu2023data}
Jiaxing Xu, Yunhan Yang, David Huang, Sophi~Shilpa Gururajapathy, Yiping Ke, Miao Qiao, Alan Wang, Haribalan Kumar, Josh McGeown, and Eryn Kwon.
\newblock Data-driven network neuroscience: On data collection and benchmark.
\newblock \emph{Advances in Neural Information Processing Systems}, 36:\penalty0 21841--21856, 2023.

\bibitem[Esteban et~al.(2019)Esteban, Markiewicz, Blair, Moodie, Isik, Erramuzpe, Kent, Goncalves, DuPre, Snyder, et~al.]{esteban2019fmriprep}
Oscar Esteban, Christopher~J Markiewicz, Ross~W Blair, Craig~A Moodie, A~Ilkay Isik, Asier Erramuzpe, James~D Kent, Mathias Goncalves, Elizabeth DuPre, Madeleine Snyder, et~al.
\newblock fmriprep: a robust preprocessing pipeline for functional mri.
\newblock \emph{Nature methods}, 16\penalty0 (1):\penalty0 111--116, 2019.

\bibitem[Jack~Jr et~al.(2008)Jack~Jr, Bernstein, Fox, Thompson, Alexander, Harvey, Borowski, Britson, L.~Whitwell, Ward, et~al.]{jack2008alzheimer}
Clifford~R Jack~Jr, Matt~A Bernstein, Nick~C Fox, Paul Thompson, Gene Alexander, Danielle Harvey, Bret Borowski, Paula~J Britson, Jennifer L.~Whitwell, Chadwick Ward, et~al.
\newblock The alzheimer's disease neuroimaging initiative (adni): Mri methods.
\newblock \emph{Journal of Magnetic Resonance Imaging: An Official Journal of the International Society for Magnetic Resonance in Medicine}, 27\penalty0 (4):\penalty0 685--691, 2008.

\bibitem[Harms et~al.(2018)Harms, Somerville, Ances, Andersson, Barch, Bastiani, Bookheimer, Brown, Buckner, Burgess, et~al.]{harms2018extending}
Michael~P Harms, Leah~H Somerville, Beau~M Ances, Jesper Andersson, Deanna~M Barch, Matteo Bastiani, Susan~Y Bookheimer, Timothy~B Brown, Randy~L Buckner, Gregory~C Burgess, et~al.
\newblock Extending the human connectome project across ages: Imaging protocols for the lifespan development and aging projects.
\newblock \emph{Neuroimage}, 183:\penalty0 972--984, 2018.

\bibitem[Wu et~al.(2022)Wu, Li, Eickhoff, Hoffstaedter, Hanke, Yeo, and Genon]{wu2022cross}
Jianxiao Wu, Jingwei Li, Simon~B Eickhoff, Felix Hoffstaedter, Michael Hanke, BT~Thomas Yeo, and Sarah Genon.
\newblock Cross-cohort replicability and generalizability of connectivity-based psychometric prediction patterns.
\newblock \emph{Neuroimage}, 262:\penalty0 119569, 2022.

\bibitem[Faskowitz et~al.(2023)Faskowitz, Moyer, Handwerker, Gonzalez-Castillo, Bandettini, Jbabdi, and Betzel]{faskowitz2023commentary}
Joshua Faskowitz, Daniel Moyer, Daniel~A Handwerker, Javier Gonzalez-Castillo, Peter~A Bandettini, Saad Jbabdi, and Richard Betzel.
\newblock Commentary on pang et al.(2023) nature.
\newblock \emph{bioRxiv}, pages 2023--07, 2023.

\bibitem[Dong et~al.(2024{\natexlab{b}})Dong, Wu, Chen, Zhang, Jin, and Zhou]{dong2024prompt}
Zijian Dong, Yilei Wu, Zijiao Chen, Yichi Zhang, Yueming Jin, and Juan~Helen Zhou.
\newblock Prompt your brain: Scaffold prompt tuning for efficient adaptation of fmri pre-trained model.
\newblock In \emph{International Conference on Medical Image Computing and Computer-Assisted Intervention}, pages 512--521. Springer, 2024{\natexlab{b}}.

\end{thebibliography}


\clearpage
\section*{NeurIPS Paper Checklist}

\begin{enumerate}

\item {\bf Claims}
    \item[] Question: Do the main claims made in the abstract and introduction accurately reflect the paper's contributions and scope?
    \item[] Answer: \answerYes{} 
    \item[] Justification: The last paragraph in introduction (Section \ref{intro}) explicitly lists the paper’s five core contributions. All claims have been verified through experimental results in Section \ref{4.3}, \ref{4.4}, and \ref{4.5}.

\item {\bf Limitations}
    \item[] Question: Does the paper discuss the limitations of the work performed by the authors?
    \item[] Answer: \answerYes{} 
    \item[] Justification: Limitations and future work are discussed in Section \ref{lim}.

\item {\bf Theory assumptions and proofs}
    \item[] Question: For each theoretical result, does the paper provide the full set of assumptions and a complete (and correct) proof?
    \item[] Answer: \answerNA{} 
    \item[] Justification: The paper does not include any theoretical result.

    \item {\bf Experimental result reproducibility}
    \item[] Question: Does the paper fully disclose all the information needed to reproduce the main experimental results of the paper to the extent that it affects the main claims and/or conclusions of the paper (regardless of whether the code and data are provided or not)?
    \item[] Answer: \answerYes{} 
    \item[] Justification: All datasets used in this work are publicly available, with detailed preprocessing information in Section \ref{4.1} and Appendix \ref{A}. The model checkpoints and source code are included in the supplementary materials, accompanied by detailed instructions in the README file.

\item {\bf Open access to data and code}
    \item[] Question: Does the paper provide open access to the data and code, with sufficient instructions to faithfully reproduce the main experimental results, as described in supplemental material?
    \item[] Answer: \answerYes{} 
    \item[] Justification: All datasets used in this work are publicly available. The model checkpoints and source code are included in the supplementary materials, accompanied by detailed instructions in the README file.

\item {\bf Experimental setting/details}
    \item[] Question: Does the paper specify all the training and test details (e.g., data splits, hyperparameters, how they were chosen, type of optimizer, etc.) necessary to understand the results?
    \item[] Answer: \answerYes{} 
    \item[] Justification: Full codebase is provided in the supplementary materials. Detailed training settings with hyperparameters are provided in Section \ref{4.1}, \ref{imp} and Appendix \ref{B}.

\item {\bf Experiment statistical significance}
    \item[] Question: Does the paper report error bars suitably and correctly defined or other appropriate information about the statistical significance of the experiments?
    \item[] Answer: \answerYes{} 
    \item[] Justification: Statistical significance is clearly labeled in the main results (Table \ref{tab1} and \ref{tab2}), with error bars in Table \ref{Ap_tab1} and \ref{Ap_tab2}. 

\item {\bf Experiments compute resources}
    \item[] Question: For each experiment, does the paper provide sufficient information on the computer resources (type of compute workers, memory, time of execution) needed to reproduce the experiments?
    \item[] Answer: \answerYes{} 
    \item[] Justification: We have provided the computer resources in Section \ref{imp}.
    
\item {\bf Code of ethics}
    \item[] Question: Does the research conducted in the paper conform, in every respect, with the NeurIPS Code of Ethics \url{https://neurips.cc/public/EthicsGuidelines}?
    \item[] Answer: \answerYes{} 
    \item[] Justification: This work conforms the NeurIPS Code of Ethics in every respect.

\item {\bf Broader impacts}
    \item[] Question: Does the paper discuss both potential positive societal impacts and negative societal impacts of the work performed?
    \item[] Answer: \answerYes{} 
    \item[] Justification: Societal impacts are discussed in Appendix \ref{C}.
    
\item {\bf Safeguards}
    \item[] Question: Does the paper describe safeguards that have been put in place for responsible release of data or models that have a high risk for misuse (e.g., pretrained language models, image generators, or scraped datasets)?
    \item[] Answer: \answerNA{} 
    \item[] Justification: The proposed research poses no such risks.

\item {\bf Licenses for existing assets}
    \item[] Question: Are the creators or original owners of assets (e.g., code, data, models), used in the paper, properly credited and are the license and terms of use explicitly mentioned and properly respected?
    \item[] Answer: \answerYes{} 
    \item[] Justification: The original papers that produced the code package or dataset are all properly cited.

\item {\bf New assets}
    \item[] Question: Are new assets introduced in the paper well documented and is the documentation provided alongside the assets?
    \item[] Answer: \answerYes{} 
    \item[] Justification: Code and model checkpoints are provided in the supplementary materials, with detailed documentation. 

\item {\bf Crowdsourcing and research with human subjects}
    \item[] Question: For crowdsourcing experiments and research with human subjects, does the paper include the full text of instructions given to participants and screenshots, if applicable, as well as details about compensation (if any)? 
    \item[] Answer: \answerNA{} 
    \item[] Justification: The datasets of human subjects used in this paper are all publicly available.

\item {\bf Institutional review board (IRB) approvals or equivalent for research with human subjects}
    \item[] Question: Does the paper describe potential risks incurred by study participants, whether such risks were disclosed to the subjects, and whether Institutional Review Board (IRB) approvals (or an equivalent approval/review based on the requirements of your country or institution) were obtained?
    \item[] Answer: \answerNA{} 
    \item[] Justification: The datasets of human subjects used in this paper are all publicly available.

\item {\bf Declaration of LLM usage}
    \item[] Question: Does the paper describe the usage of LLMs if it is an important, original, or non-standard component of the core methods in this research? Note that if the LLM is used only for writing, editing, or formatting purposes and does not impact the core methodology, scientific rigorousness, or originality of the research, declaration is not required.
    \item[] Answer: \answerNA{} 
    \item[] Justification: The core method development in this research does not involve LLMs as any important, original, or non-standard components.

\end{enumerate}

\newpage
\appendix

\section{Details of datasets}
\label{A}

In this section, we detail the datasets used for pretraining and downstream evaluation of BrainHarmonix, describing their characteristics and associated preprocessing procedures.

\begin{figure}[htbp]
    \centering
    \includegraphics[width=\columnwidth]{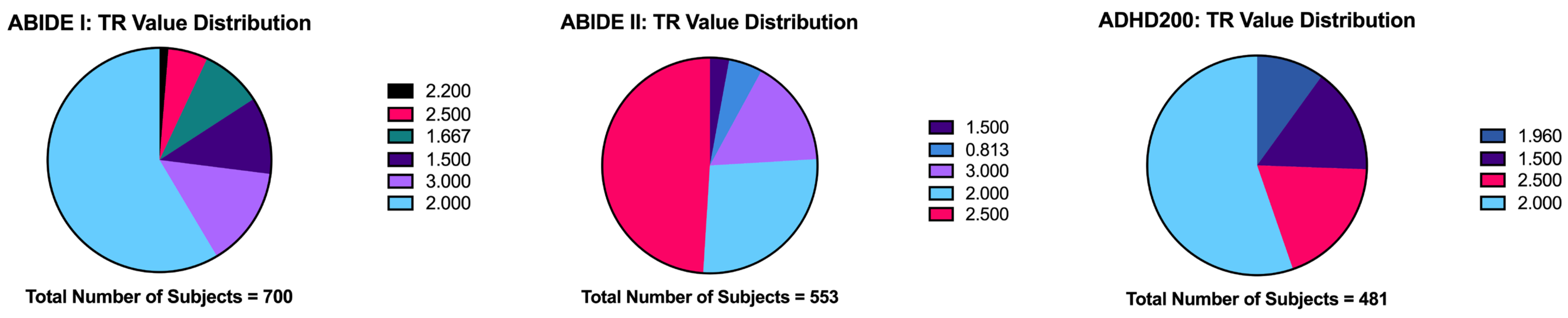}
    \caption{\textbf{TR distributions of three multi-site datasets.}}
    \label{ap1}
\end{figure}

\subsection{T1 preprocessing (shared by all datasets)}

All T1-weighted images underwent a standardized preprocessing pipeline following \cite{leonardsen2022deep}: first, images were skull-stripped using FreeSurfer \cite{segonne2004hybrid}; then reoriented to match the standard orientation defined by FMRIB Software Library (FSL) \cite{jenkinson2012fsl}; and subsequently registered to the Montreal Neurological Institute (MNI) 152 template using FSL’s linear registration tool (FLIRT) \cite{jenkinson2001global}. Finally, images were cropped to dimensions of 167 × 212 × 160 voxels, with voxel intensities normalized to a [0, 1] range.

\subsection{Adolescent Brain Cognitive Development (ABCD)}

The \href{https://abcdstudy.org/}{ABCD} Study is the largest long-term study of brain development and child health in the United States \cite{casey2018adolescent}. We curated data from 11,221 participants, each with one/two visits (baseline (aged 107-133 months) and two-year follow-up), totaling 18,139 T1-weighted images and 30,771 resting-state fMRI time series (TR=0.8s). The fMRI preprocessing pipeline was as follows: fMRI were first aligned to T1-weighted anatomical scans using boundary-based registration. Respiratory pseudo-motion artifacts were reduced by applying a band-stop filter within the 0.31-0.43 Hz range. Frames exhibiting excessive motion - defined as framewise displacement (FD) greater than 0.3 mm or voxel-wise differentiated signal variance (DVARS) exceeding 50 - were flagged. Each flagged frame, along with the preceding frame and two subsequent frames, was censored; additionally, any uncensored data segments containing fewer than five consecutive frames were also removed. Subsequently, nuisance signals - including global, white matter, ventricular signals, six head motion parameters, and their temporal derivatives - were regressed out, with coefficients computed from uncensored data. Missing data from censored frames were interpolated using the Lomb-Scargle periodogram method, after which a band-pass filter of 0.009-0.08 Hz was applied. Finally, processed data were projected onto the FreeSurfer \cite{segonne2004hybrid} fsaverage6 surface template and spatially smoothed using a Gaussian kernel with a 6 mm full-width at half-maximum (FWHM).

\subsection{UK Biobank (UKB)}

The \href{https://www.ukbiobank.ac.uk/}{UK Biobank} is a large-scale biomedical database containing in-depth health information from UK participants, with the neuroimaging component represents the largest brain imaging study \cite{bycroft2018uk,miller2016multimodal}. We curated neuroimaging data from 43,112 participants aged 44 to 83, comprising 46,455 T1-weighted MRI and 40,162 fMRI time series (TR=0.735). Following Brain-JEPA \cite{dong2024brain}, we used the preprocessed fMRI data from \cite{di2023connectomes}.

\subsection{Autism Brain Imaging Data Exchange (ABIDE-I and ABIDE-II)}

The Autism Brain Imaging Data Exchange (\href{https://fcon_1000.projects.nitrc.org/indi/abide/}{ABIDE}) is a multi-site, open-access initiative that aggregates structural and resting-state fMRI scans - alongside rich phenotypic data - from individuals with autism spectrum disorder (ASD) and matched typically developing controls to accelerate reproducible neuroimaging research. We curated neuroimaging data from two releases: ABIDE-I \cite{di2014autism}, which contains 700 participants (320 control vs. 380 ASD) with paired T1 and fMRI data collected across 20 different sites, and ABIDE-II \cite{di2017enhancing}, which includes 553 participants (230 control vs. 323 ASD) from 12 different sites. The distribution of heterogeneous TR values is shown in Figure \ref{ap1}. The fMRI preprocessing begins by de-obliquing and reorienting each fMRI run, then discarding the initial volumes before applying slice-timing correction. Head motion was corrected with FSL’s mcflirt \cite{jenkinson2012fsl}, and the time series were coregistered to each subject’s T1 image using FreeSurfer's bbregister \cite{segonne2004hybrid}. Next, nuisance signals - including global, white matter, ventricular signals, six head motion parameters, and their temporal derivatives - were regressed out and the data were despiked and band-pass filtered (0.009-0.08 Hz). Finally, the processed data were normalized to MNI space.


\begin{table}[]
  \centering
    \caption{Comparison on neurodevelopmental disorder diagnosis (standard deviations).}
  \small
  \setlength{\tabcolsep}{3pt}
  \renewcommand{\arraystretch}{1.1}
  \begin{tabularx}{\textwidth}{@{}X c c c@{\hspace{4pt}}| cc| cc| cc@{}}
    \toprule
    \multirow[b]{2}{*}{\textbf{Model}} &
      \multirow[b]{2}{*}{\textbf{Morphology}} &
      \multirow[b]{2}{*}{\textbf{Dynamics}} &
      \multirow[b]{2}{*}{\textbf{Multi-TR}} &
      \multicolumn{2}{c|}{\textbf{ABIDE-I}} &
      \multicolumn{2}{c|}{\textbf{ABIDE-II}} &
      \multicolumn{2}{c}{\textbf{ADHD-200}} \\
    \cmidrule(lr){5-6} \cmidrule(lr){7-8} \cmidrule(lr){9-10}
    & & & & \textbf{ACC}\% & \textbf{F1}\% & \textbf{ACC}\% & \textbf{F1}\% & \textbf{ACC}\% & \textbf{F1}\% \\
    \midrule
    \multicolumn{4}{@{}l}{\emph{Structure-based models}} & \multicolumn{6}{@{}l@{}}{} \\
    \midrule
    BrainMVP\textsuperscript{\scriptsize1} \cite{rui2024brainmvp}  & \cmark & \xmark & N.A.    & 0.85 & 1.68 & 4.44 & 13.11 & 10.57 & 23.24 \\
    BrainMVP\textsuperscript{\scriptsize2} \cite{rui2024brainmvp}     & \cmark & \xmark & N.A.    & 8.20 & 5.85 & 2.95 & 18.85 & 1.18 & 2.71 \\
    \textbf{BrainHarmonix-S}  & \cmark & \xmark & N.A.    & 6.32 & 5.74 & 3.79 & 4.03 & 1.56 & 4.36 \\
    \midrule
    \multicolumn{4}{@{}l}{\emph{Function-based models}} & \multicolumn{6}{@{}l@{}}{} \\
    \midrule
    BrainNetCNN \cite{kawahara2017brainnetcnn}           & \xmark & \xmark & \cmark & 1.38 & 0.75 & 1.33 & 1.90 & 0.59 & 3.70 \\
    BrainGNN \cite{li2021braingnn}              & \xmark & \xmark & \cmark & 2.67 & 3.46 & 2.19 & 2.24 & 2.05 & 3.78 \\
    BrainNetTF \cite{kan2022brain}                   & \xmark & \xmark & \cmark & 2.13 & 1.61 & 3.06 & 2.55 & 2.36 & 1.69 \\
    BrainMass \cite{yang2024brainmass}             & \xmark & \xmark & \cmark & 4.21 & 3.76 & 3.52 & 3.00 & 2.57 & 1.11 \\
    BrainLM \cite{carobrainlm}               & \xmark & \cmark & \xmark & -- & -- & -- & -- & -- & -- \\
    Brain-JEPA \cite{dong2024brain}            & \xmark & \cmark & \xmark & -- & -- & -- & -- & -- & -- \\
    \textbf{BrainHarmonix-F}  & \xmark & \cmark & \cmark & 1.38 & 2.30 & 1.32 & 1.06 & 4.60 & 2.77 \\
    \midrule
    \multicolumn{4}{@{}l}{\emph{Multimodal model}} & \multicolumn{6}{@{}l@{}}{} \\
    \midrule
    \textbf{BrainHarmonix}        & \cmark & \cmark & \cmark & 4.31 & 1.30 & 2.18 & 1.02 & 4.57 & 3.31 \\
    \bottomrule
  \end{tabularx}
  \begin{tablenotes}
   \item[*] \hspace{-.5cm}\small \textsuperscript{\scriptsize1}UniFormer \cite{li2023uniformer} as backbone; \textsuperscript{\scriptsize2}UNET3D \cite{ronneberger2015u} as backbone.
    \end{tablenotes}
  \label{Ap_tab1}
\end{table}

\begin{table}[]
  \centering
    \caption{Comparison on neurodegenerative disease classification and cognition prediction (standard deviations).}
  \small
  \setlength{\tabcolsep}{3pt}
  \renewcommand{\arraystretch}{1.1}
  \begin{tabularx}{\textwidth}{@{}X c c c@{\hspace{4pt}}| cc| cc| cc@{}}
    \toprule
    \multirow[b]{2}{*}{\textbf{Model}} &
      \multirow[b]{2}{*}{\textbf{Morphology}} &
      \multirow[b]{2}{*}{\textbf{Dynamics}} &
      \multirow[b]{2}{*}{\textbf{Multi-TR}} &
      \multicolumn{2}{c|}{\textbf{PPMI}} &
      \multicolumn{2}{c|}{\textbf{ADNI}} &
      \multicolumn{2}{c}{\textbf{HCP-A}} \\
    \cmidrule(lr){5-6} \cmidrule(lr){7-8} \cmidrule(lr){9-10}
    & & & & \textbf{ACC}\% & \textbf{F1}\% & \textbf{ACC}\% & \textbf{F1}\% & \textbf{MAE} & $\rho$ \\
    \midrule
    \multicolumn{4}{@{}l}{\emph{Structure-based models}} & \multicolumn{6}{@{}l@{}}{} \\
    \midrule
    BrainMVP\textsuperscript{\scriptsize1} \cite{rui2024brainmvp}  & \cmark & \xmark & N.A.    & 7.14 & 10.02 & 2.55 & 1.57 & 0.07 & 0.11 \\
    BrainMVP\textsuperscript{\scriptsize2} \cite{rui2024brainmvp}     & \cmark & \xmark & N.A.    & 1.34 & 1.43 & 5.25 & 15.02 & 0.11 & 0.06 \\
    \textbf{BrainHarmonix-S}  & \cmark & \xmark & N.A.    & 3.55 & 6.31 & 3.06 & 9.50 & 0.21 & 0.11 \\
    \midrule
    \multicolumn{4}{@{}l}{\emph{Function-based models}} & \multicolumn{6}{@{}l@{}}{} \\
    \midrule
    BrainNetCNN \cite{kawahara2017brainnetcnn}           & \xmark & \xmark & \cmark & 1.10 & 0.82 & 1.75 & 3.77 & 0.59 & 0.07 \\
    BrainGNN \cite{li2021braingnn}              & \xmark & \xmark & \cmark & 0.00 & 0.45 & 4.63 & 8.32 & 0.63 & 0.06 \\
    BrainNetTF \cite{kan2022brain}                   & \xmark & \xmark & \cmark & 1.35 & 1.06 & 3.03 & 2.30 & 0.23 & 0.01 \\
    BrainMass \cite{yang2024brainmass}             & \xmark & \xmark & \cmark & 1.63 & 3.28 & 1.75 & 5.81 & 0.69 & 0.06 \\
    BrainLM \cite{carobrainlm}               & \xmark & \cmark & \xmark & 2.33 & 1.56 & 5.25 & 5.08 & 0.26 & 0.03 \\
    Brain-JEPA \cite{dong2024brain}            & \xmark & \cmark & \xmark & 2.17 & 3.67 & 1.43 & 0.24 & 0.61 & 0.14 \\
    \textbf{BrainHarmonix-F}  & \xmark & \cmark & \cmark & 2.33 & 3.65 & 1.75 & 1.91 & 0.73 & 0.11 \\
    \midrule
    \multicolumn{4}{@{}l}{\emph{Multimodal model}} & \multicolumn{6}{@{}l@{}}{} \\
    \midrule
    \textbf{BrainHarmonix}        & \cmark & \cmark & \cmark & 3.55 & 6.31 & 4.63 & 3.87 & 0.56 & 0.12 \\
    \bottomrule
  \end{tabularx}
    \begin{tablenotes}
   \item[*] \hspace{-.5cm}\small \textsuperscript{\scriptsize1}UniFormer \cite{li2023uniformer} as backbone; \textsuperscript{\scriptsize2}UNET3D \cite{ronneberger2015u} as backbone.
    \end{tablenotes}
  \label{Ap_tab2}
  \label{tab:brain_models}
\end{table}

\begin{table}[]
  \caption{Pre-training settings.}
  \label{tab:Pre-training setting.}
  \renewcommand{\arraystretch}{1.1}
  \centering
\begin{tabular}{l|l}
\hline
config                   & value                        \\ \hline
\multicolumn{2}{l}{\emph{Common configs}}                      \\ \hline
$d$      & 768 \\
optimizer                & AdamW \citep{loshchilov2018decoupled}                       \\
optimizer momentum       & $\beta_1,\beta_2=0.9,0.999$                    \\
learning rate schedule   & warmup cosine schedule \citep{assran2023self}            \\ \hline
\multicolumn{2}{l}{\emph{BrainHarmonix-S \&  Harmonizer configs}}        \\ \hline
start learning rate      & 0                     \\
learning rate            & $5\times 10^{-4}$                     \\
final learning rate      & 0                     \\
weight decay schedule    & constant                     \\
weight decay             & 0.05                         \\
warmup epochs            & 40                           \\
BrainHarmonix-S patch size      & 16                           \\
$N_S$   & 1200 \\
BrainHarmonix-S total batch size & 150 $\times $ 8 GPU cards = 1200                         \\
BrainHarmonix-S training epochs & 800                          \\
BrainHarmonix total batch size & 15 $\times $ 8 GPU cards = 120                         \\
BrainHarmonix training epochs   & 50                           \\ \hline
\multicolumn{2}{l}{\emph{BrainHarmonix-F configs}}                    \\ \hline
$J$  & 200 \cite{pang2023geometric} \\
patch size, $k*$              & 48                           \\
$\tau$ & $48\times 0.735=35.28$ seconds \\
max number of tokens   &  18 \\
$N_F$   & $400\times 18=7200$ \\
start learning rate      & $2.5\times 10^{-6}$                     \\
learning rate            & $5.7\times 10^{-5}$                     \\
final learning rate      & $1 \times 10^{-6}$                     \\
weight decay schedule    & cosine weight decay schedule \citep{assran2023self} \\
weight decay             & 0.05                         \\
final weight decay       & 0.4                          \\
EMA momentum schedule    & linear \citep{assran2023self}                      \\
EMA start momentum       & 0.996                        \\
EMA final momentum       & 1                            \\
total batch size         & 64 $\times$ 8 GPU Cards = 512                          \\
warmup epochs            & 10                           \\
training epochs          & 100                          \\  \hline
\end{tabular}
\end{table}

\subsection{Attention Deficit Hyperactivity Disorder (ADHD-200)}

The \href{https://fcon_1000.projects.nitrc.org/indi/adhd200/}{ADHD-200} dataset is a multi-site, open-access repository of structural and resting-state fMRI scans with accompanying phenotypic measures from children and adolescents with attention-deficit/hyperactivity disorder (ADHD) and matched typically developing controls. We curated neuroimaging data from 481 participants (292 control vs. 189 ADHD) with paired T1 and fMRI data collected across 6 different sites. The distribution of heterogeneous TR values is shown in Figure \ref{ap1}. The ADHD-200 dataset underwent the same preprocessing procedure as the ABIDE datasets.

\subsection{Parkinson's Progression Markers Initiative (PPMI)}

The \href{https://www.ppmi-info.org/}{PPMI} is a longitudinal, multi-center, open-access dataset combining imaging, biospecimens, and detailed clinical assessments from Parkinson’s disease (PD) patients, prodromal cohorts, and healthy controls to accelerate biomarker discovery and disease-progression research \cite{marek2011parkinson}. We utilized the open benchmark repository \cite{xu2023data} preprocessed by fMRIPrep \cite{esteban2019fmriprep}, which contains data from 195 participants (15 control, 14 SWEDD, 53 Prodromal, and 113 PD patients).

\subsection{Alzheimer’s Disease Neuroimaging Initiative (ADNI)}

The \href{https://adni.loni.usc.edu/data-samples/adni-data/}{ADNI} is a longitudinal, multi-site study providing open-access neuroimaging, biomarker, genetic, and clinical data from cognitively normal (CN), mild cognitive impairment (MCI), and Alzheimer’s disease (AD) participants to advance early diagnosis and therapeutic research \cite{jack2008alzheimer}. We curated neuroimaging data from 164 participants (83 CN vs. 81 MCI) with paired T1 and fMRI data. The preprocessing procedure is the same as the ABIDE datasets.

\subsection{Lifespan Human Connectome Project Aging (HCP-A)}

The \href{https://www.humanconnectome.org/study/hcp-lifespan-aging}{HCP-A} dataset is a Lifespan Human Connectome Project release that provides multimodal MRI and rich behavioral assessments from adults to elucidate brain connectivity changes across healthy aging \cite{harms2018extending}. The resting-state fMRI data in MNI152 space underwent ICA-FIX denoising. We then performed nuisance regression to control for 24 motion parameters, white matter signal, CSF signal, and their temporal derivatives following \cite{wu2022cross}.

\section{Additional implementation details}
\label{B}
The default optimization settings for pretraining are detailed in Table \ref{tab:Pre-training setting.}. We initialized all transformer blocks using the Xavier uniform method, as described in \citep{he2022masked}. For downstream adaptation, the default setting follows MAE \cite{he2022masked}, except for using AdmaW for linear probing.

\section{Additional analysis \& discussion}
\label{C}

\subsection{Synthetic testing of TAPE}

\begin{table}[h]
\centering
\caption{Performance comparison on HCP-A test sets}
\begin{tabular}{lcc}
\toprule
& \textbf{MAE} & \textbf{Correlation} \\
\midrule
Original test set & 6.56 & 0.42 \\
Synthetic test set & 6.69 & 0.39 \\
\bottomrule
\end{tabular}
\label{tape}
\end{table}

To further evaluate TAPE’s effectiveness, we tested on both the original HCP-A test set and version with samples randomly downsampled by factors of 1 and 2 (equal probability, leading to TR values 1.6, 2.4). The comparable performance across conditions demonstrates TAPE's robustness (Table \ref{tape}).

\subsection{Performance improvement: parameter count v.s. multimodal integration}

\begin{table}[h]
\centering
\caption{Performance comparison across model sizes (accuracy\%)}
\begin{tabular}{lccccc}
\toprule
& \textbf{Single Modality (fMRI)} & \textbf{Concat} & \textbf{22M} & \textbf{86M} & \textbf{307M} \\
\midrule
ABIDE-II & 62.90 & 63.19 & 64.06 & 66.67 & 66.95 \\
ADHD-200 & 67.69 & 68.36 & 69.39 & 70.09 & 70.40 \\
\bottomrule
\end{tabular}
\label{gain}
\end{table}

We performed additional comparisons to better illustrate the performance gains from incorporating multiple modalities (2nd \& 3rd columns in the Table \ref{gain}  containing the results regarding accuracy (\%)) and from introducing the harmonizer module for fusion (4th-6th columns). Specifically, we concatenated the embeddings from the frozen T1 and fMRI encoders and passed them through a trainable linear layer for the classification task. Furthermore, we conducted experiments using harmonizers of different sizes, ranging from 22M parameters to 307M parameters (4th-6th columns). The results above clearly demonstrate the performance improvements achieved both by adding modalities and by scaling the harmonizer module.

\subsection{Ablation of ABCD dataset}

\begin{table}[h]
\centering
\caption{Model performance comparison on different training data (MCI classification on ADNI)}
\begin{tabular}{lcc}
\toprule
\textbf{Model} & \textbf{ACC (\%)} & \textbf{F1 Score (\%)} \\
\midrule
Brain-JEPA using UKB only & 59.60 & 60.78 \\
BrainHarmonix-F using UKB only & 60.67 & 63.34 \\
BrainHarmonix-F using both UKB \& ABCD & 61.62 & 64.80 \\
\bottomrule
\end{tabular}
\label{abcd}
\end{table}

We conducted an ablation on ADNI for MCI classification by pretraining BrainHarmonix-F without ABCD data (Table \ref{abcd}). We found it still outperformed the original Brain-JEPA based on the same UKB dataset. 

\subsection{Ablation studies on ADNI}

\begin{table}[h]
\centering
\caption{Ablation on ADNI (accuracy\%)}
\begin{tabular}{lccc}
\toprule
& \textbf{BrainHarmonix w/o pre-alignment} & \textbf{BrainHarmonix-F} & \textbf{BrainHarmonix} \\
\midrule
with DA & 61.35 & 61.62 & 64.65 \\
w/o DA & 60.07 & 60.11 & 62.94 \\
\bottomrule
\end{tabular}
\begin{tablenotes}
   \item[*] \hspace{-.1cm}\small DA: data augmentation.
    \end{tablenotes}
\label{adni}
\end{table}

We additionally performed ablation studies on ADNI (Table \ref{adni} regarding accuracy (\%)), where we observed a similar trend and performance pattern, reinforcing the effectiveness of our proposed model design.

\subsection{Generalization to Asian clinical cohorts}

\begin{table}[h]
\centering
\caption{Performance comparison on MACC}
\begin{tabular}{lcc}
\toprule
& \textbf{ACC (\%)} & \textbf{F1 (\%)} \\
\midrule
BrainMVP & 65.83 & 53.64 \\
BrainHarmonix-S & 67.68 & 56.67 \\
BrainNetCNN & 57.57 & 52.00 \\
BrainGNN & 62.61 & 40.57 \\
BrainNetTF & 63.03 & 57.57 \\
BrainMass & 64.65 & 57.93 \\
BrainLM & 63.64 & 54.03 \\
Brain-JEPA & 66.67 & 59.18 \\
BrainHarmonix-F & 68.69 & 62.50 \\
\textbf{BrainHarmonix} & \textbf{74.75*} & \textbf{65.57*} \\
\bottomrule
\end{tabular}
\label{macc}
\end{table}

We extended our evaluation to an Asian clinical cohort collected by Memory, Ageing and Cognition Center (\href{https://www.ukbiobank.ac.uk/}{MACC}), thereby assessing generalizability to non-Western populations and in real-world clinical scenarios. Specifically, we performed an additional task - classification of amyloid-positive/negative participants, which holds significant clinical value for AD prognosis and intervention. As shown in the Table \ref{macc}, BrainHarmonix achieved state-of-the-art performance in this clinically relevant, in-house setting, underscoring its robustness and cross-population generalizability. 

\subsection{Evaluation across data portions}

\begin{table}[h]
\centering
\caption{Accuracy vs data portion (fine-tuning)}
\begin{tabular}{llccccc}
\toprule
& \textbf{Portion (\%)} & \textbf{20\%} & \textbf{40\%} & \textbf{60\%} & \textbf{80\%} & \textbf{100\%} \\
\midrule
ABIDE-II & Accuracy (\%) & 55.94 & 56.52 & 60.87 & 63.77 & 66.67 \\
ADHD-200 & Accuracy (\%) & 57.14 & 62.72 & 67.44 & 69.39 & 70.09 \\
\bottomrule
\end{tabular}
\label{portion}
\end{table}

\begin{table}[h]
\centering
\caption{Accuracy vs data portion (pretraining)}
\begin{tabular}{lccccc}
\toprule
\textbf{Pretrain Portion} & \textbf{20\%} & \textbf{40\%} & \textbf{60\%} & \textbf{80\%} & \textbf{100\%} \\
\midrule
ABIDE-II & 59.12 & 62.90 & 64.35 & 65.21 & 66.67 \\
ADHD-200 & 64.96 & 65.64 & 67.01 & 68.37 & 70.09 \\
\bottomrule
\end{tabular}
\label{portion_pretrain}
\end{table}

We conducted additional analyses by scaling the fine-tuning dataset using increasing proportions (20\%, 40\%, 60\%, 80\%, and 100\%). The results regarding accuracy (\%) are shown in the Table \ref{portion}. Our results demonstrate a clear and consistent scaling of performance with increasing data portions. Notably, compared with prior leading baseline BrainMass 59.35\% on ABIDE-II and 65.99\% on ADHD, BrainHarmonix achieves state-of-the-art performance even when fine-tuned on only 80\% of the dataset, highlighting the efficiency and effectiveness of our pretrained representations.

On the other hand, we investigated the effect of using different portions of the pretraining dataset. Specifically, we applied identical sampling proportions to both the UKB and ABCD datasets for pretraining. The corresponding results regarding accuracy (\%) are reported in the Table \ref{portion_pretrain}. We observe that the model’s performance improves as the portion of the pretraining dataset increases.

\subsection{Scaling with increasing token numbers}

\begin{table}[h]
\centering
\caption{Scaling with increasing token numbers (accuracy\%)}
\begin{tabular}{llcccccc}
\toprule
\textbf{Dataset} & \textbf{Method} & \textbf{32} & \textbf{64} & \textbf{128} & \textbf{256} & \textbf{512} & \textbf{1024} \\
\midrule
ABIDE-II & Finetune & 62.61 & 65.21 & 66.67 & 66.96 & 67.53 & 66.96 \\
& Linear Probe & 61.45 & 61.45 & 61.74 & 62.03 & 62.32 & 62.89 \\
ADHD-200 & Finetune & 67.69 & 69.05 & 70.09 & 70.41 & 70.41 & 70.75 \\
& Linear Probe & 66.33 & 67.69 & 68.37 & 68.71 & 68.37 & 69.05 \\
\bottomrule
\end{tabular}
\label{scaling}
\end{table}

For completeness, we have included results with 512 and 1024 tokens as references in addtion to the results in the main content. As shown in the Table \ref{scaling}, the accuracy (\%) remains relatively stable beyond 256 tokens, confirming our initial observation.

\subsection{Efficiency evaluation}

We included the pretraining time (on 8 NVIDIA H100 GPUs (80GB)), as well as finetuning time (on 1 H100 GPU) and inference time (on 1 H100 GPU) on ABIDE-II, corresponding to different model sizes (token counts) to provide a more comprehensive view of the computational cost in the Table \ref{eff}. Larger model or more token counts lead to longer computing time. 

\begin{table}[!h]
\centering
\caption{Training and inference time comparison across model sizes (token numbers) on ABIDE-II}
\resizebox{\textwidth}{!}{%
\begin{tabular}{lcccccccl}
\toprule
& \textbf{22M (128)} & \textbf{307M (128)} & \textbf{86M (32)} & \textbf{86M (64)} & \textbf{86M (128)} & \textbf{86M (256)} & \textbf{86M (512)} & \textbf{86M (1024)} \\
\midrule
Pretraining Time & 5h 10m & 17h 9m & 9h 20m & 9h 26m & 9h 37m & 9h 45m & 10h 23m & 11h 11m \\
FT Training Time & 0h 21m 52s & 1h 07m 28s & 0h 25m 33s & 0h 26m 54s & 0h 27m 41s & 0h 29m 54s & 0h 30m 17s & 0h 31m 54s \\
Inference Time & 5.02s & 7.19s & 5.90s & 5.89s & 6.36s & 5.77s & 6.11s & 7.47s \\
\bottomrule
\end{tabular}
}
\label{eff}
\begin{tablenotes}
   \item[*] \hspace{-.1cm}\small FT: fine tuning.
    \end{tablenotes}
\end{table}

\subsection{Discussion on dynamic time warping (DTW)}

Dynamic time warping (DTW) is an algorithm that measures the similarity between two temporal sequences, or time series, that may vary in speed or timing. It assumes a meaningful temporal correspondence between sequences. However, in the context of resting-state fMRI, there is no ground truth temporal alignment across individuals, as each subject’s brain dynamics evolve independently and asynchronously. Therefore, applying DTW across different scans would impose artificial temporal correspondences not supported by the data.

\subsection{Discussion on geometric harmonics in neuroimaging community}

There are critiques to \cite{pang2023geometric}, which focus on the paper’s claim that geometric harmonics, by themselves, can serve as a “winner-take-all” solution for brain dynamics reconstruction, thereby diminishing the role of the structural connectome \cite{faskowitz2023commentary}. However, the critiques do not affect the validity of our geometric pre-alignment. The harmonics in our work are only used to provide geometry-aware positioning, we make no claim that they can fully explain/reconstruct brain dynamics. On the other hand, the harmonics are averaged with large-scale functional gradients, so it does not have a winner-take-all basis. Future work can explore how structural connectome and other biological principles can be encoded into the model and whether they can further improve brain representation learning and generalizability.

\subsection{Potential failure mode}

In our current evaluations, one notable case where BrainHarmonix underperforms is on the ADHD-200. Its F1 is slightly lower than BrainHarmonix-F. This is likely due to motion artifacts in T1, as ADHD patients exhibit increased head motion during MRI acquisition. Such motion introduces noise and negatively affects structural data quality, potentially reducing multimodal fusion performance. Future work will explore methods to improve robustness against data-quality issues. On the other hand, although we have demonstrated data scaling effects, model performance under low-sample and few-shot learning scenarios remains an area for improvement. Future studies may address few-shot adaptation through approaches such as parameter-efficient fine-tuning or prompt-based tuning \cite{dong2024prompt}.

\section{Broader Impact}
\label{D}

The integration and compression of multimodal neuroimaging signals could not only reduces storage and computational demands but also lays the groundwork for deployment on resource-constrained platforms. Coupled with its capabilities for various downstream tasks, it may accelerates exploratory analyses, supports richer biomarker discovery, and drives improvements in diagnosis, prognosis, and personalized treatment planning. Moreover, its capacity to harmonize data across different scanners and acquisition protocols could enhance reproducibility and deepens our understanding of large‑scale neuroimaging in both health and disease. However, these powerful capabilities also bring ethical responsibilities. Protecting patient confidentiality and ensuring data integrity are essential—deployments must include rigorous de‑identification procedures and secure data pipelines.


\end{document}